\documentclass{article}
\usepackage[utf8]{inputenc}
\usepackage{hyperref}       
\usepackage{amsmath}
\usepackage{graphicx}
\usepackage{bm} 
\usepackage{color}
\usepackage[normalem]{ulem}
\usepackage{fancyhdr}

\usepackage{arxiv}
\newcommand{\gam}{{\gamma}}

\newcommand{\eps}{{\epsilon}}

\newcommand{\sig}{{\sigma}}

\newcommand{\Ome}{{\Omega}}
\usepackage[linesnumbered, ruled, vlined]{algorithm2e}

\newcommand{\rmd}{\ensuremath{\mathrm{d}}}
\newcommand{\dint}{\mbox{\, \rmd}}

\newcommand{\Dt}{\Delta t}

\newcommand{\Yc}{Y_\mathrm{c}}
\newcommand{\lc}{\ell_\mathrm{c}}

\newcommand{\dotd}{\dot{d}}

\newcommand{\abs}[1]{\lvert #1 \rvert}

\newcommand{\dbar}{\overline{d}}

\newcommand{\dt}{\Delta t}

\newcommand{\up}{u_{\mathrm{p}}}
\newcommand{\vp}{v_{\mathrm{p}}}

\newcommand{\Dn}{D}

\newcommand{\piu}{\pi^\mathrm{u}}
\newcommand{\pil}{\pi^\mathrm{\ell}}

\newcommand{\sigc}{\sig_{\mathrm{c}}}

\newcommand{\wc}{w_{\mathrm{c}}}

\newcommand{\Lip}{\mathrm{L}}

\newcommand{\helem}{h_\mathrm{e}}
\newcommand{\gd}{g}
\newcommand{\lamc}{\lambda_\mathrm{c}}
\newcommand{\epsdoti}{\dot{\epsilon}_0}
\newcommand{\darg}{\tilde{d}}
\newcommand{\uarg}{\tilde{u}}
\newcommand{\tnew}{t}

\newcommand{\astfootnote}[1]{%
\let\oldthefootnote=\thefootnote%
\setcounter{footnote}{0}%
\renewcommand{\thefootnote}{\fnsymbol{footnote}}%
\footnote{#1}%
\let\thefootnote=\oldthefootnote%
}

\title{Fragmentation analysis of a bar with the Lip-field approach}
\author{Nicolas Mo\"{e}s, Beno\^{i}t L\'{e} and Andrew Stershic}
\date{\today\\ \vspace{10pt} SAND2022-0325 O}

\author{Nicolas Mo\"es \\
	\'Ecole Centrale de Nantes\\
	GeM Institute, UMR CNRS 6183 \\
	Institut Universitaire de France (IUF) \\
	1 rue de la No\"{e}, 44321 Nantes, France \\
	\texttt{nicolas.moes@ec-nantes.fr} \\
	\And
    Beno\^it L\'e  \astfootnote{Corresponding author} \\
	\'Ecole Centrale de Nantes\\
	GeM Institute, UMR CNRS 6183 \\
	1 rue de la No\"{e}, 44321 Nantes, France \\
	\texttt{benoit.le@ec-nantes.fr} \\
	\And
	Andrew Stershic \\
    Sandia National Laboratories, California\\
	7011 East Avenue, Livermore, CA 94550, USA \\
	\texttt{ajsters@sandia.gov} \\
	}
		
\hypersetup{
pdftitle={Fragmentation analysis of a bar with the Lip-field approach},
pdfauthor={Nicolas Mo\"{e}s, Beno\^{i}t L\'{e} and, Andrew Stershic},
pdfkeywords={Fragmentation, Damage, Fracture, Dynamic, Lipschitz, Lip-field},
}

\begin{document}
\maketitle

\begin{abstract}
   
    The Lip-field approach was introduced in \cite{Moes2021} as a new way to regularize softening material models. It was tested in 1D quasistatic in \cite{Moes2021} and 2D quasistatic in \cite{Chevaugeon2021}: this paper extends it to 1D dynamics, on the challenging problem of dynamic fragmentation. The Lip-field approach formulates the mechanical problem to be solved as an optimization problem, where the incremental potential to be minimized is the non-regularized one. Spurious localization is prevented by imposing a Lipschitz constraint on the damage field. The displacement and damage field at each time step are obtained by a staggered algorithm, that is the displacement field is computed for a fixed damage field, then the damage field is computed for a fixed displacement field. Indeed, these two problems are convex, which is not the case of the global problem where the displacement and damage fields are sought at the same time. The incremental potential is obtained by equivalence with a cohesive zone model, which makes material parameters calibration simple. A non-regularized local damage equivalent to a cohesive zone model is also proposed. It is used as a reference for the Lip-field approach, without the need to implement displacement jumps. 
    These approaches are applied to the brittle fragmentation of a 1D bar with randomly perturbed material properties to accelerate spatial convergence. Both explicit and implicit dynamic implementations are compared. Favorable comparison to several analytical, numerical and experimental references serves to validate the modeling approach. 

\end{abstract}


\section{Introduction}
\label{sec:introduction}

%
%
%
%

Fragmentation is a challenging application for failure modeling approaches, as failure is pervasive through the domain. It is indeed a multiscale problem, balancing the localization processes that lead to fracture surfaces with a significant interactivity between the multiple failure points. Accordingly, these competing physics lead to a material response varies significantly with loading rate; with quasistatic and low-velocity loadings, a small number of dominant cracks is sufficient to relieve the structure; in contrast, strongly dynamic loadings lead to rate-dependent multiple fragmentation
\cite{Denoual2000}.

Much study on fragmentation has been motivated by military application, with foundational studies performed by Mott \cite{Mott1947}, supplemented by a number of experimental studies (e.g. \cite{Grady1982}), analytical models (e.g. \cite{Glenn1986,Drugan2001}), and numerical models (e.g. \cite{Miller1999,Pandolfi1999,Zhou2006}). The original motivation persists (e.g. \cite{Pearson1990,Tonge2016}), and the topic has additional relevance in such diverse fields such as ice-sheet morphology (e.g. \cite{Selvadurai2009,Aastrom2019}), geophysics (e.g. \cite{Bishop2016,Hu2020}), and nuclear reactor fuels (e.g. \cite{Jiang2020}).

Fracture of softening materials is physically characterized by a process zone ahead of the cracks tips where micro-cracking nucleate and coalesce to give macro-cracks. Taking into account the characteristic length of this process zone is a key point in numerical modelling of softening materials.

A first and very popular model to consider this characteristic length is the cohesive zone model (CZM) \cite{Dugdale1960,Baranblatt1961,hilleborg1976,Park2011}. It consists in modeling the macro-cracks by displacement jumps, with cohesive forces acting on the crack lips over a certain distance before the crack tips. These cohesive zones are usually introduced at the boundaries between finite elements, which make them very dependent on the mesh orientation. Some approaches like the eXtended finite element method (X-FEM) \cite{Moes2002} alleviates the mesh dependency by allowing the introduction of cohesive cracks inside the finite elements, however modelling complex crack patterns like branching or coalescence remains challenging from a geometry perspective.

Other means of discrete fracture representation that have been used to model fragmentation include various particle methods, such as: smooth particle hydrodynamics \cite{Pramanik2015}, material point method \cite{Li2015}, discrete element method \cite{Kun1996}, and peridynamics \cite{Lai2015}.

An alternative class of numerical failure modeling is continuum damage models.
These models represent the influence of micro-cracks by means of some internal damage variable \cite{Kachanov1958,chaboche1988}. With this approach complex crack morphology (e.g. branching, networks, fragmentation) is handled natively, even if the exact position of macro-cracks is not available anymore. The failure of the material is represented by a softening constitutive model; however, local damage models suffer from the well-known problem of spurious mesh dependency \cite{Bazant1976}, that is a dissipated energy which tends to zero when refining finite element meshes.

These models need a length scale $\lc$ to recover well-posedness. This problem has been widely studied, see for instance higher-order gradient models \cite{Cosserat1909,Chambon1998} or regularization of internal variables \cite{Pijaudier1987,Lasry1988,Peerlings1996,Jirasek1998,giry2011b}. Alternatively, a local damage model can be regularized by introducing a characteristic time instead of a characteristic length \cite{Needleman1988,Dube1996,Allix1997,Suffis2003}. 
Delay damage models and nonlocal integral models are compared in \cite{Desmorat2010}; it seems that even if both models can regularize local damage models, their validity domain depends on the strain rates. A recent assessment of the delay damage model 
in limit cases may also be found in \cite{Zghal2020}.

Another approach of recent interest is the phase-field fracture model \cite{Karma2001,Miehe2010}, which derives from a smoothing of fracture mechanics.
Based on the variational approach to fracture \cite{Francfort1998}, it uses a smoothed representation of the macro-cracks to get a process zone with a finite thickness and can deal with complex crack topologies. It was initially developed for brittle fracture, but more recent works have extended it to cohesive fracture \cite{Verhoosel2013,Wu2018}, enabling application to ductile fracture \cite{Talamini2021}. The phase-field approach has also been applied to dynamic brittle fracture \cite{Borden2012,Ren2019,Geelen2019,Fischer2019}. 

The Thick Level Set (TLS) approach introduced by \cite{moes2011,Stolz2012b} defines the damage variable as a user-defined function of a level-set function corresponding to the distance to the damage front (boundary between the damaged and undamaged material). In this framework, the iso-$\lc$ of this level-set function gives the crack position, which can be enriched to introduce displacement jumps \cite{Bernard2012}. Complex crack patterns can be captured using specific element cutting algorithms \cite{salzman2016}. The TLS has been applied to dynamic fracture \cite{Moreau2015} and fragmentation \cite{Stershic2017}. 

A new regularization method, called the Lip-field approach, was introduced in \cite{Moes2021}. It is based on formulating the mechanical problem to be solved as an optimization problem, where the unknowns are the displacement field and the damage variable (with eventually other internal variables like plastic deformation for instance). The expression of the potential to be minimized is exactly the same as the non-regularized problem, however a Lipschitz constraint is applied on the damage variable to keep its gradient bounded. This approach was tested in 1D and 2D on quasistatic simulations \cite{Moes2021,Chevaugeon2021}, with damage and softening plasticity (1D case) models. 
Compared to the TLS approach, the Lip-field approach does not require the level-set technology and yields 
a convex optimization on the damage variable. Also, 
as for the phase-field approach the Lip-field 
solution proceeds with a staggered scheme in which each step is rather straightforward to implement in the finite element framework (a Python Lip-field implementation is provided 
in \cite{Chevaugeon2021}). 


In this paper, the Lip-field is extended to 1D dynamics on the aforementioned fragmentation problem.
The paper is organized as follows. Section \ref{sec:problemEquations} describes the mechanical formulation of the 1D fragmentation problem. Section \ref{sec:lipfield} focuses on the Lip-field regularization. Section \ref{sec:CZMEquivalence} explains how to get a Lip-field model equivalent to a linear CZM. It also gives a simplified implementation of a CZM which can be used as comparison. Details about the numerical resolution are given in Section \ref{sec:numericalResolution}. Some numerical results are presented in Section \ref{sec:numericalExamples}. Section \ref{sec:conclusion} concludes the paper. 

\section{The 1D fragmentation problem}
\label{sec:problemEquations}

We consider the deformation with respect to the time variable $t$ of a 1D bar $\Ome = [0; L]$, represented by its elongation $u(x,t)$ for $(x,t) \in \Ome \times [0;T_{\text{max}}]$. In the framework of small-strains assumption, the uniaxial strain $\epsilon$ writes
\begin{equation}
 \epsilon = u_{,x}
\end{equation}
We look for the bar displacement $u \in \mathcal{U}$ such that
\begin{equation}
 \label{eq:equilibrium}
 \int_{0}^{L} \sigma \eps(u^*) \dint x = \int_{0}^{L} \rho \ddot{u} u^* \dint x, \quad \forall u^* \in \mathcal{U^*}
\end{equation}
where $\sigma$ is the uniaxial stress and $\rho$ the density. $\mathcal{U}(t)$ is the space of the kinematically admissible solutions
\begin{equation}
 \mathcal{U}(t) = \left\lbrace u \in H^1(\Ome) : u(0,t) = 0, u(L,t) = \epsdoti L t \right\rbrace
\end{equation}
and $\mathcal{U^*}$ the test functions space
\begin{equation}
 \mathcal{U^*} = \left\lbrace u \in H^1(\Ome) : u^*(0) = 0, u^*(L) = 0 \right\rbrace
\end{equation}
$\epsdoti$ is the initial, uniform strain rate that is imposed on the bar as initial condition:
\begin{align}
 u(x,0) & =  0, \quad  \forall x \in \Ome \\
 \dot{u}(x,0) & =   \epsdoti x, \quad  \forall x \in \Ome
\end{align}

The bar is made of a brittle material characterized by a free energy 
\begin{equation}
    \varphi(\epsilon, d) = \frac{1}{2} g(d) E \epsilon^2 
\end{equation}
where $d$ is a damage variable and $E$ the Young modulus. $g(d)$ is a decreasing function such that $g(0) = 1$ (undamaged material) and $g(1) = 0$ (fully damaged material). The state equations are
\begin{equation}
    \sigma = \frac{\partial \varphi}{\partial \epsilon} =  E g(d) \epsilon, \quad Y =  - \frac{\partial \varphi}{\partial d} = - \frac{1}{2} g'(d) E \epsilon^2 
\end{equation}
where the prime indicates a derivative with respect to $d$.
The evolution laws are chosen as 
\begin{equation}
    \dotd \geq 0, \quad Y - \Yc H(d) \leq 0, \quad (Y - \Yc H(d)) \dotd = 0
\end{equation}
where $\Yc$ is the critical energy release rate and $H$ is an increasing function such that $H(0) = 1$, called the ``softening function'' in what follows. Note that the energy release rate is symmetrical in tension/compression. Asymmetry did not seem necessary to obtain good results as it will be shown in Section \ref{sec:numericalExamples}, but could be easily implemented if needed, see \cite{Chevaugeon2021}.

Solving the fragmentation problem described by the above equations requires to determine, at each time $t_n$, the displacement field $u_n$ and the damage field $d_n$. In what follows, we will assume that $(u_n,d_n)$ are known, and we seek to compute $(u_{n+1},d_{n+1})$ at time $t_{n+1} = t_n + \dt$ (the ``$n+1$'' indexes are hereafter dropped for simplicity). It can be formulated as an optimization problem:  
\begin{subequations}
\label{eq:localProblem}
\begin{align}
    \label{eq:localProblemFirst}
    (u,d) & = \arg \min_{\uarg \in U, \darg \in \Dn} F(\uarg, \darg) \\
    \label{eq:localProblemSecond}
    F(u, d) & = \int_0^L \frac{1}{2}  \rho \frac{( u - \up)^2}{\beta \Delta t^2} + \varphi(\epsilon(u), d) + \Yc h(d) \dint x 
\end{align}
\end{subequations}
where
\begin{equation}
 h(d)  = \int_0^{d} H(\darg) \dint \darg
 \label{eq:hprimitive}
\end{equation}
and
\begin{align}
     U & = \left\lbrace u \in H^1(\Ome) : u(0) = 0, u(L) = \epsdoti L t_{n+1} \right\rbrace \\
\label{eq:DnSpace}
 \Dn  & = \{  d \in L^\infty(\Ome): d_n \leq  d \leq 1     \}  
\end{align}
Note that the definition of $\Dn$ automatically ensures the irreversibility of the evolution of $d$. The predictor $\up$ is defined by 
\begin{align}
    \up & = u_n + \Dt v_n + \frac{\Dt^2}{2} (1-2\beta) a_n \label{eq:dispPrediction} \\
    \vp & = v_n + (1-\gam) \Dt a_n \label{eq:speedPrediction}
\end{align}
with $v$ the velocity and $a$ the acceleration. The numerical parameters $\beta$ and $\gam$ determine the type of numerical integration (explicit or implicit). Then, $a$ is computed (see detail in Section~\ref{sec:numericalResolution}) and gives the final values for $u$ and $v$
\begin{align}
    u & = \up + \beta \Dt^2 a \label{eq:dispCorrection} \\
    v & = \vp + \gamma \Dt a  \label{eq:speedCorrection}
\end{align}

The choice $\beta=0$ corresponds to the explicit dynamics (central difference method). The new displacement is found by $u = \up$ and thus will not depend on the coming value of damage. The case $\beta = 1/4$, $\gamma = 1/2$ corresponds to the implicit Newmark scheme (undamped trapezoidal rule). Note that for the explicit scheme, it is common to consider a lumped mass matrix, whereas for the implicit scheme, an exact mass matrix integration is performed. 

Solving \eqref{eq:localProblem} in space $\Dn$ is an ill-posed problem, which leads to solutions exhibiting spurious mesh dependency. Therefore, the Lip-field approach \cite{Moes2021} introduces an alternative problem  
\begin{equation}
\label{eq:lipProblem}
    (u,d)  = \arg \min_{\uarg \in U, \darg \in \Dn \cup \Lip} F(\uarg, \darg) 
\end{equation}
where
\begin{equation} 
\label{eq:lipSpace}
	\Lip  = \{ d \in L^{\infty}([0,L]): \abs{d(x)-d(y)} \leq \abs{x-y}/\ell, \quad \forall x, y \in [0,L] \}
\end{equation}
The only difference between problems \eqref{eq:localProblem} and \eqref{eq:lipProblem} is the space where the solution $d$ is computed: by enforcing that $d$ must belong to $\Lip$, we prevent the gradient of $d$ to go to infinity, thus avoiding spurious localization. In practice, this problem is not convex with respect to $(u,d)$, therefore, it is solved with a staggered iterative process\footnote{Here, we distinguish between timesteps and iterations, with the former annotated as subscripts (e.g. $d_n$) and the latter as superscripts ($d^{k}$).}:

\begin{subequations}
\label{eq:alternateMinimization}
\begin{align} 
 u^{k+1}  & = \arg \min_{\uarg \in U} F(\uarg, d^k) \label{eq:alternateMinimizationU} \\
 d^{k+1}  & = \arg \min_{\darg \in \Dn \cup \Lip} F(u^{k+1}, \darg) \label{eq:alternateMinimizationD}
\end{align} 
\end{subequations}

where $\dbar$ represents a suitable predictor of the damage field.
Problem \eqref{eq:alternateMinimizationU} is trivially convex with respect to $u$ for a fixed $d$, whereas problem \eqref{eq:alternateMinimizationD} is convex provided that $h(d)$ is convex and $g(d)$ convex, both bounded from below, so that it has a unique solution for a given $u$. 
 

%
%
%
%

\section{Lip-field regularization}
\label{sec:lipfield}

\begin{figure}
 \centering
 \includegraphics[width=12cm]{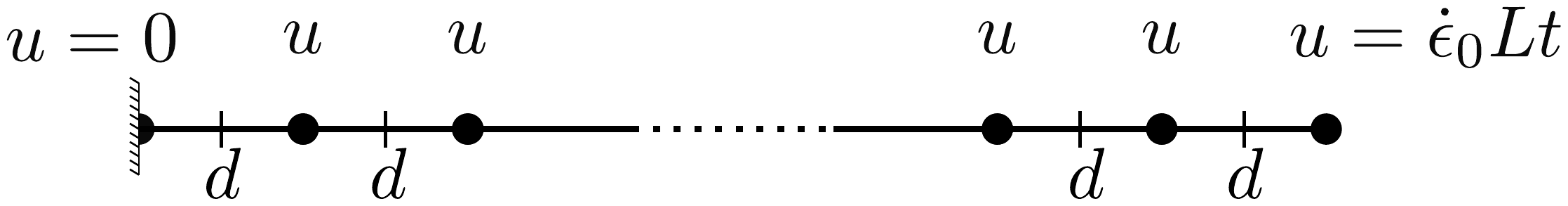}
 \caption{Discretization of a 1D bar. Displacements are stored at nodes, and damage variables at the integration points, located at center of each element.}
\label{fig:lipfield1DDispAndDamageDiscretization}
\end{figure}

In this section, we focus on the minimization problem with respect to $d$ given by equation \eqref{eq:alternateMinimizationD}, finding a solution $d$ which is in space $\Lip$. Detailed explanation and demonstration can be found in \cite{Moes2021}. In 1D, the domain is discretized with $N_e$ linear finite elements of equal size $\helem$, where the displacement $u$ is stored at each finite element node, while the damage $d$ is stored at the element integration points (element centroids), as illustrated on Figure \ref{fig:lipfield1DDispAndDamageDiscretization}.

\begin{figure}
 \centering
 \includegraphics[height=6cm]{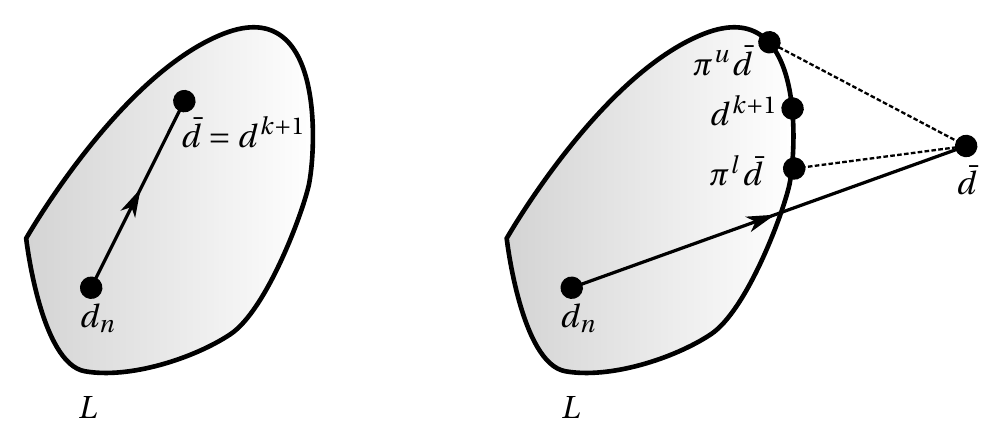}
 \caption{Sketch of computation of local update $d^{k+1}$ from $d_n$ \cite{Moes2021}. If $\dbar$ is in $\Lip$, then $d^{k+1} = \dbar$. Otherwise, $\pil \dbar$ and $\pil \dbar$ give bounds for the projection of $\dbar$ on $\Lip$.}
\label{fig:lipProj}
\end{figure}

In this case, the Lipschitz constraint on the damage variable simply writes
\begin{subequations}
 \begin{align}
 d_i - d_{i+1} - \helem / \ell & \leq 0, \quad i = 1, ... N_e-1 \\
  d_i - d_{i-1} - \helem / \ell & \leq 0, \quad i = 2, ... N_e
\end{align}
\end{subequations}
The above condition can be provided to any optimizer to solve \eqref{eq:alternateMinimizationD}. This problem can be solved in the entire computation domain, however \cite{Moes2021} proposed an efficient way to reduce the size of the domain where problem \eqref{eq:alternateMinimizationD} needs to be solved, which is used in this paper. The first step is to solve \eqref{eq:alternateMinimizationD} without the Lipschitz constraint:

\begin{equation}
\label{eq:damageLipPrediction}
 \dbar   = \arg \min_{\darg \in \Dn} F(u^{k+1}, \darg)
\end{equation}
This provides a prediction $\dbar$ of the damage field. If it satisfies the Lipschitz constraint, then $d^{k+1} = \dbar$ (see Figure \ref{fig:lipProj}). Otherwise, it was shown by \cite{Moes2021} that $d^{k+1}$ can be bounded by a lower projection $\pil$ and an upper projection $\piu$
\begin{subequations}
\label{eq:lipProjections}
 \begin{align}
 \pil \dbar(x) = \min_{y \in \Ome} \left( \dbar(y) + \frac{1}{\ell} |x-y| \right) \\
 \piu \dbar(x) = \min_{y \in \Ome} \left( \dbar(y) - \frac{1}{\ell} |x-y| \right)
\end{align}
\end{subequations}
such that
\begin{equation}
 d_n \leq \pil \dbar \leq \dbar \leq \piu \dbar \leq 1
\end{equation}
and 
\begin{equation}
 \pil \dbar \leq d^{k+1} \leq \piu \dbar
\end{equation}
An important consequence is that 
\begin{equation}
 \pil \dbar(x) = \piu \dbar(x) \Rightarrow d^{k+1}(x) = \dbar(x)
\end{equation}
We can make use of this property to solve \eqref{eq:alternateMinimizationD} only where $\pil \dbar(x) \neq \piu \dbar(x)$, thus reducing the size of the problem to be solved.

\begin{figure}
 \centering
 \includegraphics[height=6cm]{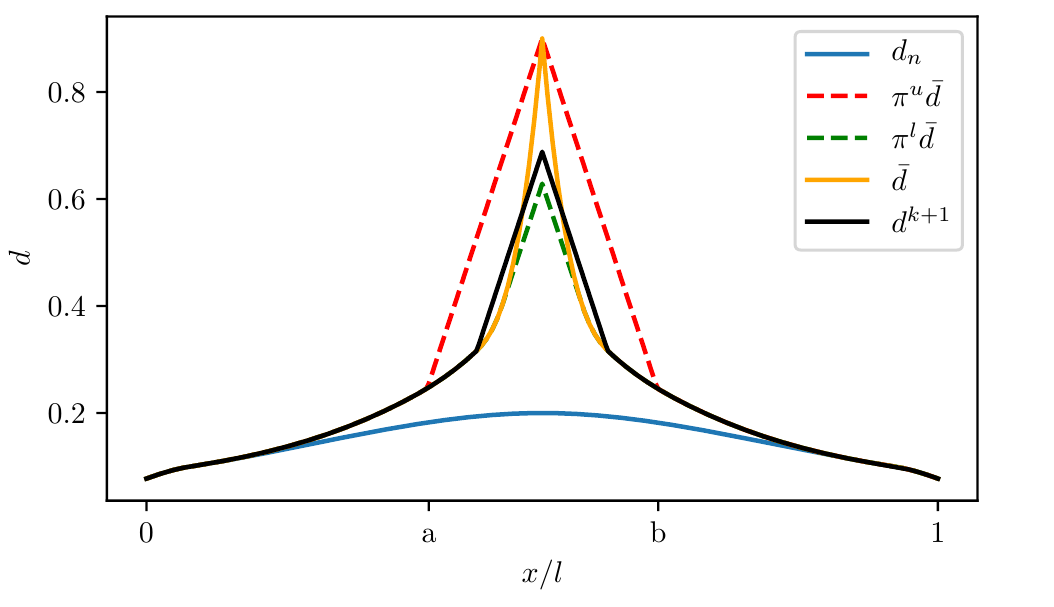}
 \caption{Illustration of the properties of the upper and lower projections \cite{Moes2021}. Outside the interval $[a,b]$, the projections are equal and thus $d^{k+1} = \dbar$. Inside $[a,b]$, we have $\pil \dbar \leq d^{k+1} \leq \piu \dbar$}
\label{fig:lipBounds}
\end{figure}

\section{Equivalence with a cohesive zone model (CZM)}
\label{sec:CZMEquivalence}

\begin{figure}
 \centering
 \includegraphics[height=5cm]{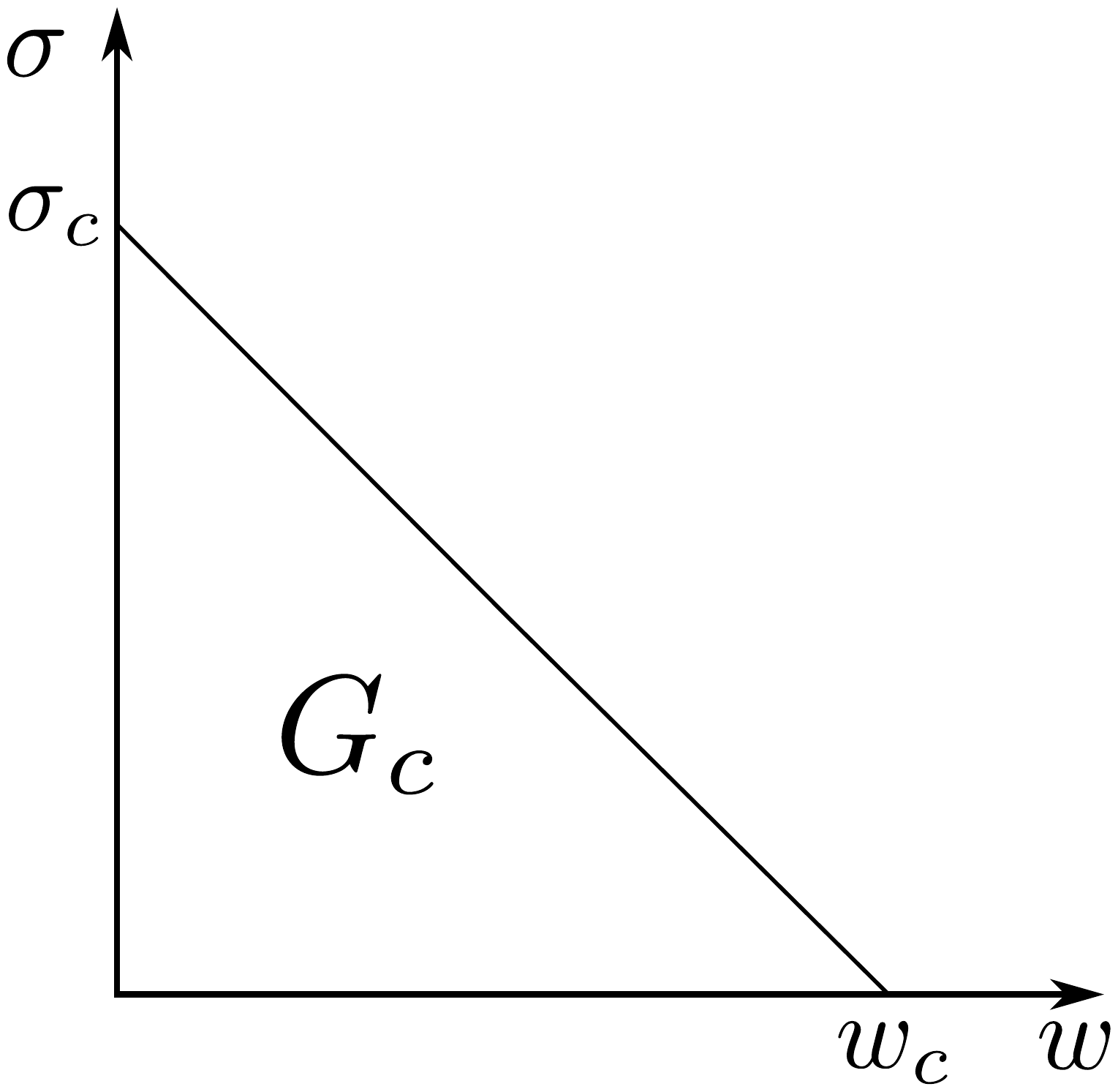}
 \caption{Linear traction-separation law for cohesive zone}
\label{fig:cohesiveLinearLaw}
\end{figure}

In this section, we discuss the expression of the softening function, and in particular its primitive $h$, introduced in equation \eqref{eq:hprimitive}. In \cite{ParrillaGomez2015}, it was shown on a 1D case that for a particular choice of $h$, it was possible to mimic the behavior of a linear cohesive zone model (CZM) (see Figure \ref{fig:cohesiveLinearLaw}). The same reasoning is used in this paper.

Considering for $g$ the following expression:
\begin{equation}
\label{eq:damageFunction}
 g(d) = (1-d)^2
\end{equation}
The expression of $h$ which allows to mimic a linear CZM is
\begin{equation}
\label{eq:hLip}
 h(d) = \frac{2d - d^2}{(1-d+\lambda d^2)^2}
\end{equation}
where 
\begin{equation}
 \lambda = \frac{2 \Yc \ell}{G_c}
\end{equation}
where $G_c$ is the material toughness. Note that $h$ is convex provided that $\lambda \leq 1/2$. Also, in the framework of the CZM equivalence, the material must start breaking when the stress is equal to the critical stress $\sigc$, which imposes
\begin{equation}
 \Yc = \frac{\sigc^2}{2E}
 \label{eq:Yc}
\end{equation}
In what follows, we will simply call ``Lip-field'' the Lip-field model associated with the above $g(d)$ and $h(d)$ functions.


We have explained how to get a Lip-field model equivalent to a linear CZM\footnote{{\color{black} We note that the Lip-field model derived herein is equivalent to a linear CZM in the sense that it features the same parameters as a linear CZM ($\{G_c,\sigma_c\}$) and produces the same crack-opening behavior in 1D quasistatics for a single flaw. This is analogous to the derivation for TLS-CZM equivalence derived in  \cite{ParrillaGomez2015}. For general application, such as the present focus on 1D dynamics with multiple opening flaws, the behavior is expected to be similar, but not to correspond absolutely.}}, so it would be logical to compare the results obtained with a true linear CZM.
However, implementing a CZM is tedious, as it requires to implement displacement jumps, eventually by introducing extra degrees of freedom.  This is why we chose in this paper an alternate way, inspired by the so-called crack-band model from \cite{Bazant1982,bazant1983}, which consists in using a damage model depending on the finite element size $\helem$. We consider a purely local damage model, so that $(u,d)$ is solution of the following problem

\begin{subequations}
\label{eq:localCZMProblem}
\begin{align}
    (u,d) & = \arg \min_{\uarg \in U, \darg \in \Dn} F_{\text{CZM}}(\uarg, \darg) \\
    F_{\text{CZM}}(u, d) & = \int_0^L \frac{1}{2}  \rho \frac{(u - \up)^2}{\beta \Delta t^2} + \varphi(\epsilon(u), d) + \Yc h_{\text{CZM}}(d) \dint x 
\end{align}
\end{subequations}
where
\begin{equation}
\label{eq:hczm}
 h_{\text{CZM}}(d)  = \frac{1}{(1-\lamc)} \left( \frac{1}{(1-\lamc) \gd(d) + \lamc} - 1 \right)
\end{equation}
with
\begin{equation}
 \lamc = \frac{\sigc \helem}{E \wc}
\end{equation}

The first difference with the Lip-field problem described by equations \eqref{eq:lipProblem} is that $d$ does not need to belong to $\Lip$ anymore. The second difference is the softening function $h_{\text{CZM}}$, which is computed so that the behavior of a single finite element of size $\helem$ has the same behavior as a finite element modeled with a CZM (the details of the derivation of $h_{\text{CZM}}$ can be found in the appendix). 


Note that in problem \eqref{eq:localCZMProblem}, $d$ does not belong to $\Lip$, so its gradient may eventually go to infinity. However, because of the dependency of $h_{\text{CZM}}(d)$ on the finite element size $\helem$, the dissipated energy does not go to zero when the mesh is refined.

In what follows, we will simply call ``CZM'' the local damage model associated with the above $h_{\text{CZM}}(d)$ function.

%
%
%
%
%
%
%
%
%
%

\section{Numerical resolution}
\label{sec:numericalResolution}

The displacement, speed, and acceleration fields are discretized using classical linear interpolation functions:
\begin{align}
 u(x,t) & = \sum_{i \in N} u_i(t) N_i(x) \\
 v(x,t) & = \sum_{i \in N} v_i(t) N_i(x) \\
 a(x,t) & = \sum_{i \in N} a_i(t) N_i(x)
\end{align}
where $N$ is the set of the finite element nodes, as illustrated on Figure \ref{fig:lipfield1DDispAndDamageDiscretization}, $u_i$, $v_i$, $a_i$ are the degrees of freedom associated to node $i$ and $N_i$ the corresponding interpolation function. In what follows, we will note $(\bm{U}_n,\bm{V}_n,\bm{A}_n)$ the vectors containing the degrees of freedom of $(u_n,v_n,a_n)$. They follow the same naming convention (predicted values, iterate, values at $\tnew$, etc.). By injecting this discretization in equation \eqref{eq:equilibrium}, we get the following matrix system:
\begin{equation}
\label{eq:matrixEquilibrium}
 [\bm{M}] \{\bm{A}_n\} +  [\bm{K}(d_n)] \{\bm{U}_n\} = \{\bm{0}\}
\end{equation}
where
\begin{align}
 M_{ij} & =  \int_{0}^{L} \rho N_i N_j \dint x \quad \forall (i,j) \in N \times N \\
 K_{ij}(d_n) & = \int_{0}^{L} E g(d_n) \rho N_{i,x} N_{j,x} \dint x \quad \forall (i,j) \in N \times N
\end{align}
Now we discuss how to solve \eqref{eq:matrixEquilibrium}. Firstly, the Dirichlet boundary conditions at $x = 0$ and $x = L$ are imposed by setting the first and last degrees of freedom of $\{\bm{U}_n\}$ to $0$ and $\epsdoti L t_{n+1}$ respectively, and by setting the corresponding values of $\{\bm{A}_n\}$ to $0$. Then, as stated in Section \ref{sec:problemEquations}, the computation of $u$ depends on the values of the parameters $\beta$ and $\gamma$.


\paragraph{ Explicit dynamic : $\beta = 0$ and $\gamma = 1/2$.}

 If $\beta = 0$, one can see from equation \eqref{eq:localProblemSecond} that the only solution of the problem is 
 \begin{equation}
  \label{eq:dispCorrectionExplicit}
  \{\bm{U}\} = \{\bm{U}_p\} =  \{\bm{U}_n\} + \Dt \{\bm{V}_n\} + \frac{\Dt^2}{2} \{\bm{A}_n\}
 \end{equation}
The predicted nodal velocity vector is
 \begin{equation}
 \label{eq:velocityPredictorExplicit}
  \{\bm{V}_p\} = \{\bm{V}_n\} + \frac{\Dt}{2} \{\bm{A}_n\}
 \end{equation}

Knowing $\{\bm{U}\}$, problem \eqref{eq:alternateMinimizationD} can be solved to compute $d$. Then, the nodal acceleration vector $\{\bm{A}\}$ is computed by solving \begin{equation}
\label{eq:matrixEquilibriumExplicit}
 [\bm{\tilde{M}}] \{\bm{A}\}  = -[\bm{K}(d)] \{\bm{U}\}
\end{equation}
where $[\bm{\tilde{M}}]$ is the lumped version of operator $[\bm{M}]$. Finally, nodal velocity vector $\{\bm{V}\}$ is obtained by equation \eqref{eq:speedCorrection}:
 \begin{equation}
 \label{eq:velocityCorrectorExplicit}
  \{\bm{V}\} = \{\bm{V}_p\} + \frac{\Dt}{2} \{\bm{A}\}
 \end{equation}
 
 The explicit dynamic resolution is summarized in Algorithm \ref{algo:explicit}.

\begin{algorithm}
 \caption{Explicit dynamic resolution}
 \label{algo:explicit}
  Initialization: $n = 0$, $\{\bm{U}_0\}$, $\{\bm{V}_0\}$, $\{\bm{A}_0\}$, $d_0$   \;
  \While{$n < n_{\text{max}}$} 
  {
  Compute nodal displacement $\{\bm{U}\}$ using \eqref{eq:dispCorrectionExplicit} and predicted nodal velocity $\{\bm{V}_p\}$ using \eqref{eq:velocityPredictorExplicit} \;
  Solve \eqref{eq:damageLipPrediction} to compute damage prediction $\dbar$ \;
  Compute projections $\piu \dbar$ and $\pil \dbar$ using \eqref{eq:lipProjections} \;
  Solve \eqref{eq:alternateMinimizationD} where $\piu \dbar \neq \pil \dbar $ \ to find $d$ \;
  Compute nodal acceleration $\{\bm{A}\}$ by solving \eqref{eq:matrixEquilibriumExplicit} \;
  Compute nodal velocity using $\{\bm{V}\}$ \eqref{eq:velocityCorrectorExplicit} \;
  $n \leftarrow n+1$ \;
  }
   
\end{algorithm}

\paragraph{Implicit dynamic : $\beta = 1/2$ and $\gamma = 1/4$.}

The first step is to compute the vector of nodal displacement prediction $\{\bm{U}_p\}$ using \eqref{eq:dispPrediction}
 \begin{equation}
  \label{eq:dispPredictorImplicit}
  \{\bm{U}_p\} = \{\bm{U}_n\} + \Dt \{\bm{V}_n\} + \frac{\Dt^2}{2} (1- 2 \beta) \{\bm{A}_n\}
 \end{equation}

Then, unlike the explicit dynamic case, iterations are necessary to compute $\{\bm{U}\}$ and $d$. By injecting the expression of $\up$ in \eqref{eq:alternateMinimizationU}, we get the linear system that needs to be solved to find $\{ \bm{U}^{k+1}\}$:
\begin{equation}
\label{eq:matrixEquilibriumImplicit}
 \left(\beta \Dt^2 [\bm{K}(d^{k})]+ [\bm{M}] \right) \{\bm{U}^{k+1}\} =  [\bm{M}] \{\bm{U}_p\} 
\end{equation}
Problem \eqref{eq:alternateMinimizationD} is solved to compute $d^{k}$. Alternate resolution of \eqref{eq:matrixEquilibriumImplicit} and \eqref{eq:alternateMinimizationD} are performed until convergence. Finally, the nodal acceleration vector is obtained by replacing in \eqref{eq:dispCorrection} the expression of $\up$ from equation \eqref{eq:dispPrediction}:
\begin{equation}
\label{eq:accelerationImplicit}
 \{\bm{A}\} = \frac{1}{\beta \Dt^2} \left(\{\bm{U}\} - \{\bm{U}_p\}\right) - \frac{1}{\beta \Dt} \{\bm{V}_n\} - \left(\frac{1}{2 \beta}-1\right) \{\bm{A}_n\} 
\end{equation}
and the nodal velocity vector by equation \eqref{eq:speedCorrection}
\begin{equation}
\label{eq:velocityCorrectorImplicit}
 \{\bm{V}\} =  \{\bm{V}_n\} + \frac{1}{2} (1-\gamma) \Dt \{\bm{A}_n\} + \gamma \Dt \{\bm{A}\} 
\end{equation}

 The implicit dynamic resolution is summarized in Algorithm \ref{algo:implicit}.


\begin{algorithm}
 \caption{Implicit dynamic resolution}
 \label{algo:implicit}
  Initialization: $n = 0$, $\{\bm{U}_0\}$, $\{\bm{V}_0\}$, $\{\bm{A}_0\}$, $d_0$   \;
  \While{$n < n_{\text{max}}$} 
  {
  Compute nodal displacement prediction $\{\bm{U}_p\}$ using \eqref{eq:dispPredictorImplicit} \;
 $ k = 0 $ \;
 $ d_0 = d_{n} $ \; 
    \While{$k < k_{\text{max}}$ \textbf{and} $err_u > tol_u$ \textbf{and} $err_d > tol_d$  } 
  {
  
  Compute nodal displacement $\{\bm{U}^{k+1}\}$ by solving \eqref{eq:matrixEquilibriumImplicit} \;
  Solve \eqref{eq:damageLipPrediction} to compute damage prediction $\dbar^{k+1}$ \;
  Compute projections $\piu \dbar^{k+1}$ and $\pil \dbar^{k+1}$ using \eqref{eq:lipProjections} \;
  Solve \eqref{eq:alternateMinimizationD} where $\piu \dbar^{k+1} \neq \pil \dbar^{k+1} $ to find $d^{k+1}$ \; 
  Compute errors :
  \begin{equation}
   err_u = \left| \frac{ \{ \bm{U}^{k+1}\} - \{\bm{U}^{k}\}}{\{\bm{U}^{k+1}\}-\{\bm{U}_{n}\}} \right|, \quad 
   err_d = \left| \frac{d^{k+1} - d^{k}}{d^{k+1}-d_{n}} \right|
  \end{equation}
  $k \leftarrow k+1$ \;
  } 
  Compute nodal acceleration $\{\bm{A}\}$ using \eqref{eq:accelerationImplicit} and nodal velocity $\{\bm{V}\}$ using \eqref{eq:velocityCorrectorImplicit} \;
  $n \leftarrow n+1$ \;
  }
   
\end{algorithm}

\section{Numerical examples}
\label{sec:numericalExamples}

%

In this section, we present the results obtained with the Lip-field and the CZM. The considered example is a 1D bar of length $L = 2 \cdot 10^{-3}$ m and cross-sectional area $A = 2 \cdot 10^{-7}$ m$^2$ from \cite{Miller1999,Drugan2001} made of dense alumina. The numerical values of the material parameters are given in Table \ref{table:materialProperties}. Following the reasoning of \cite{Stershic2017}, the Lip-field regularization length is set to $\ell = 2.21 \cdot 10^{-6}$ m, which represents half the size of the expected fragment size from Drugan's analysis \cite{Drugan2001} for the highest strain rate of $\epsdoti = 7.5 \cdot 10^{6}$ s$^{-1}$ tested in \cite{Stershic2017}. Both will be tested with explicit and implicit dynamic resolution, with a time step of $\Dt = \displaystyle 0.99 \frac{\helem}{c}$ with $c = \displaystyle \sqrt{{E}/{\rho}}$ the elastic wave speed.

A stochastic approach is employed in order to hasten mesh convergence of global quantities in the otherwise-uniform fragmentation simulations, following the observations of Molinari \emph{et al.} \cite{Molinari2007} (random mesh spacing) and Stershic \emph{et al.} \cite{Stershic2017} (random material properties).
Here, the Young's modulus follows a Weibull distribution, which can be obtained as explained in \cite{Stershic2017} by:
\begin{equation}
 E(r) = E_{0} (-\ln(r))^{1/m} + E_{min}
\end{equation}
Where $r \sim \mathcal{U}(0,1)$ is a uniformely distributed random variable. The value of $m$ is set to 2, which corresponds to a Rayleigh distribution. The values of $E_{0}$ and $E_{min}$ are computed to get an average value equal to $E$ and a coefficient of variation equal to $CV = 0.01$, which gives
\begin{equation}
 E_{min} = E (1.0 - 1.9130584 \cdot CV),  \quad E_{0} =  2.1586552 \cdot E \cdot CV
\end{equation}

\begin{table}
\centering
\begin{tabular}{llcc}
  \hline
  Properties  & Units     &  Symbol    &  Value   \\ \hline
  Density & kg$\cdot$m$^{-3}$ &  $\rho$    &  $3.9 \cdot 10^3$    \\
  Young's modulus & Pa &  $E$    &  $ 610 \cdot 10^9$    \\
  Fracture toughness & N/m & $G_c$ & $83.13$ \\
  Critical energy release rate & Pa & $\Yc$ & $820 \cdot 10^3$ \\
  Critical tensile stress & Pa & $\sigc$ & $1.0 \cdot 10^9$ \\
  \hline
 \end{tabular}
 \caption{Material properties for numerical example}
 \label{table:materialProperties}
\end{table}

\begin{figure}
\centering
\begin{tabular}{cc}
 \includegraphics[width=7.5cm]{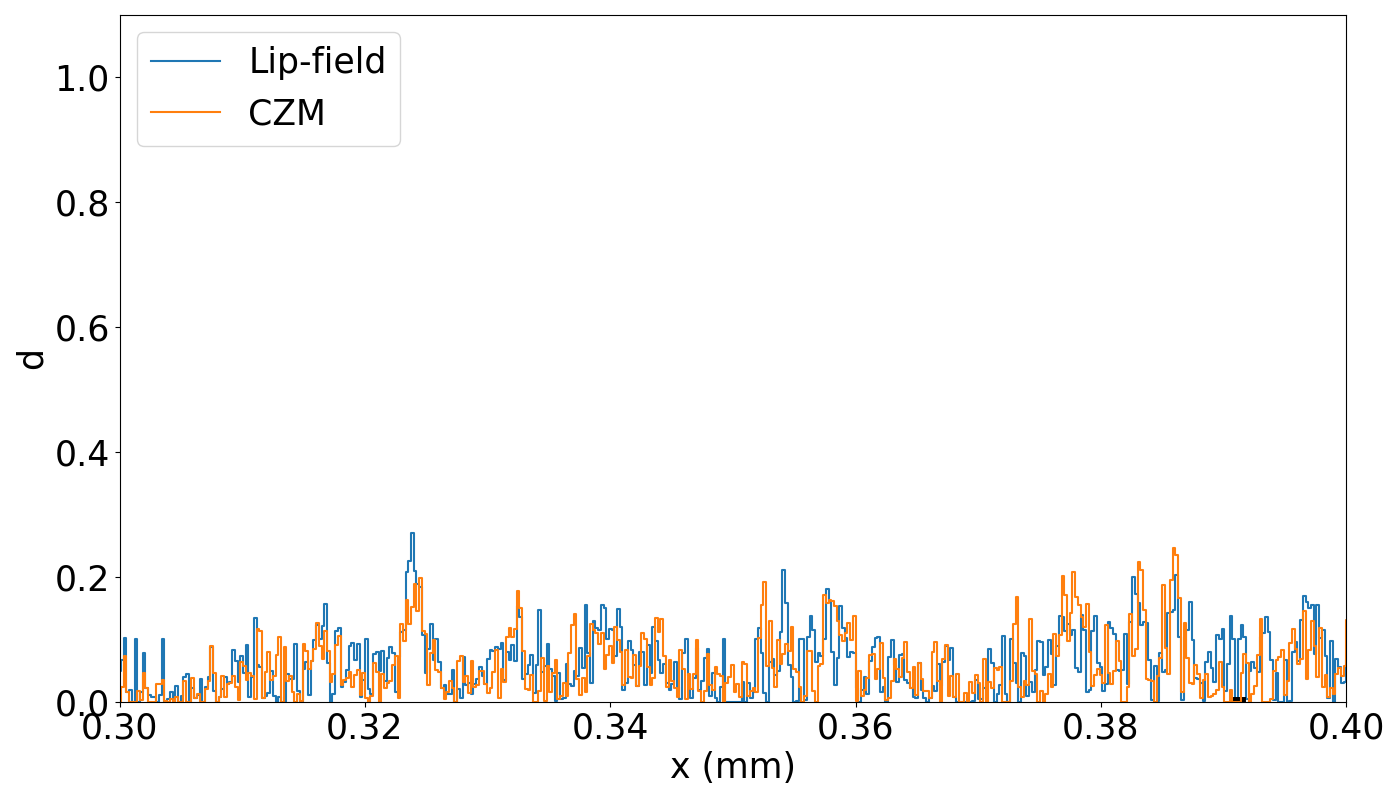}
& 
\includegraphics[width=7.5cm]{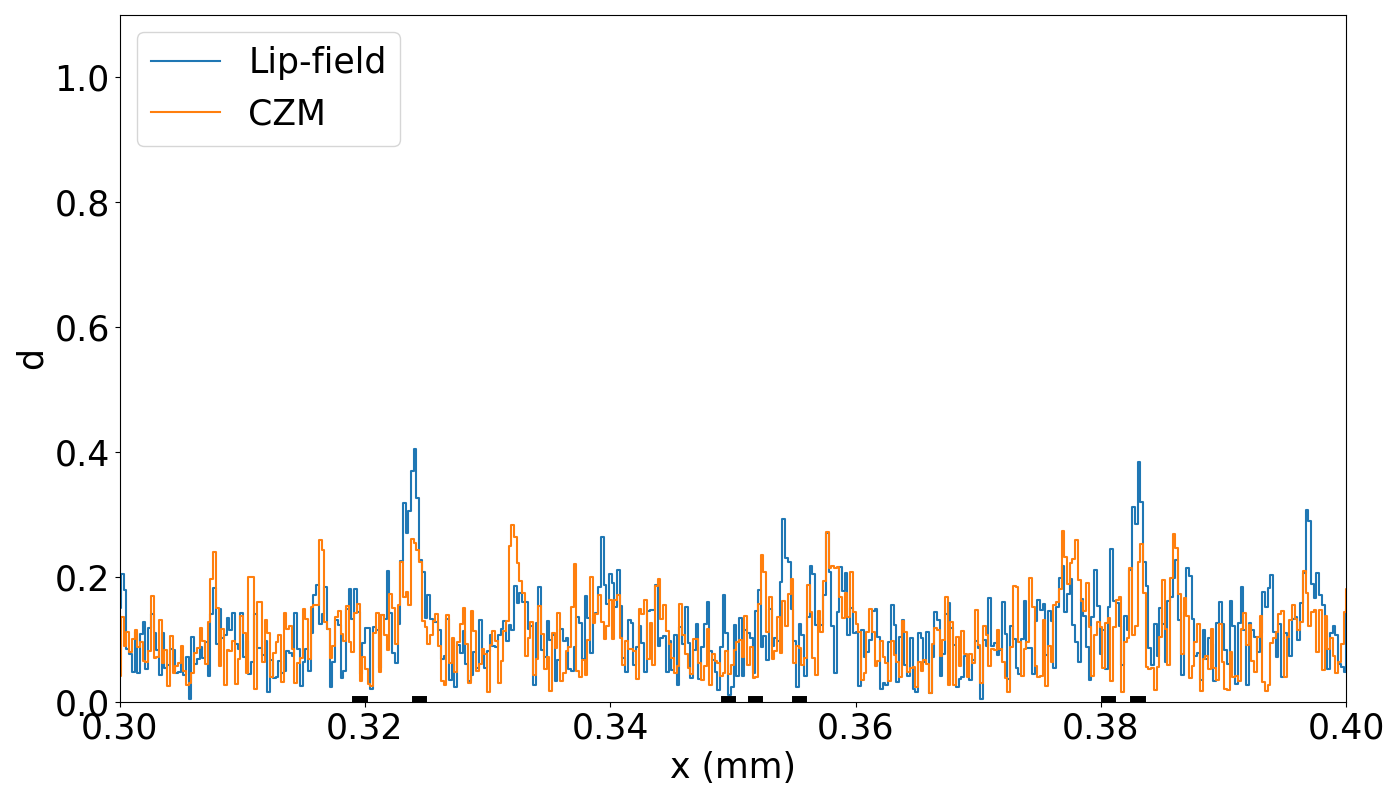}
 \\
 (a) t = 12.364 ns & (b) t = 20.136 ns \\
 \includegraphics[width=7.5cm]{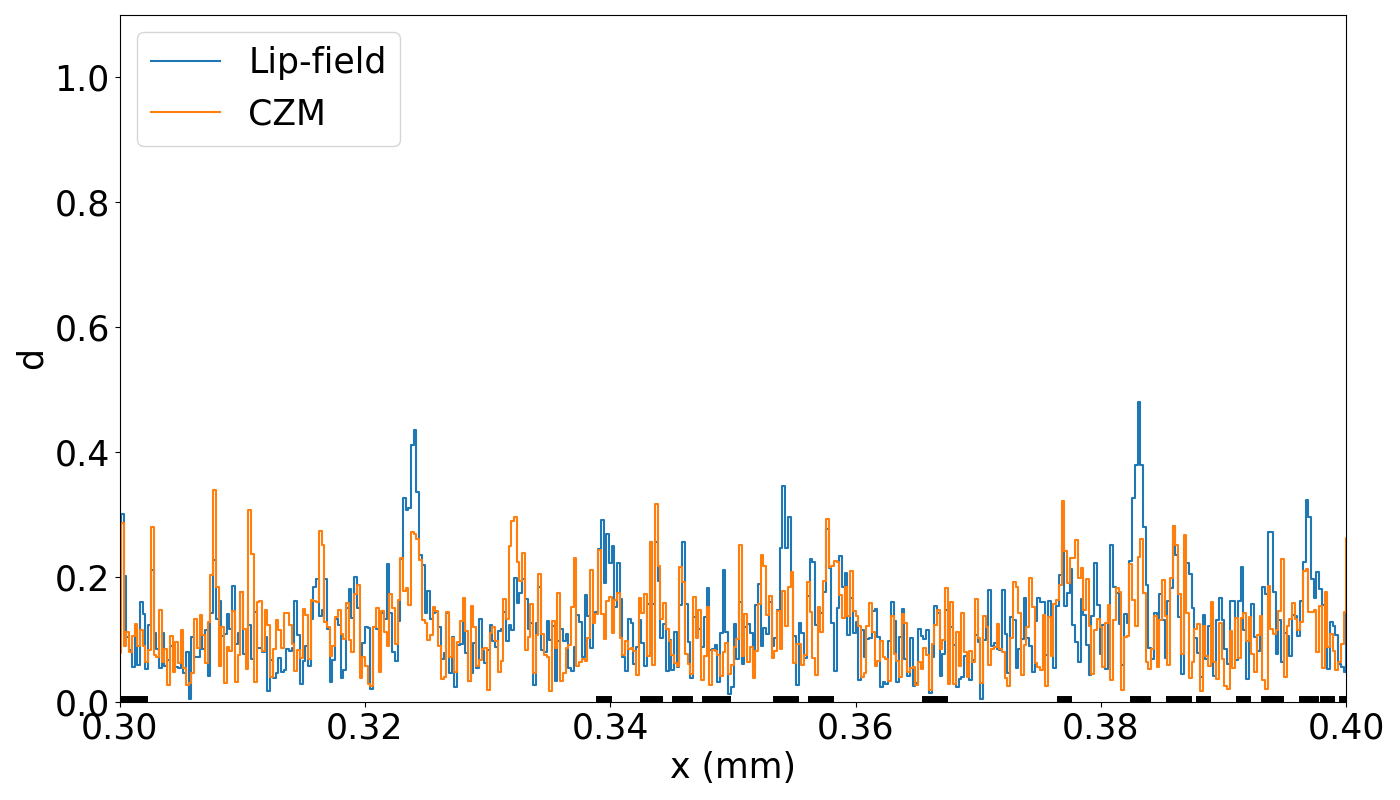}
& 
\includegraphics[width=7.5cm]{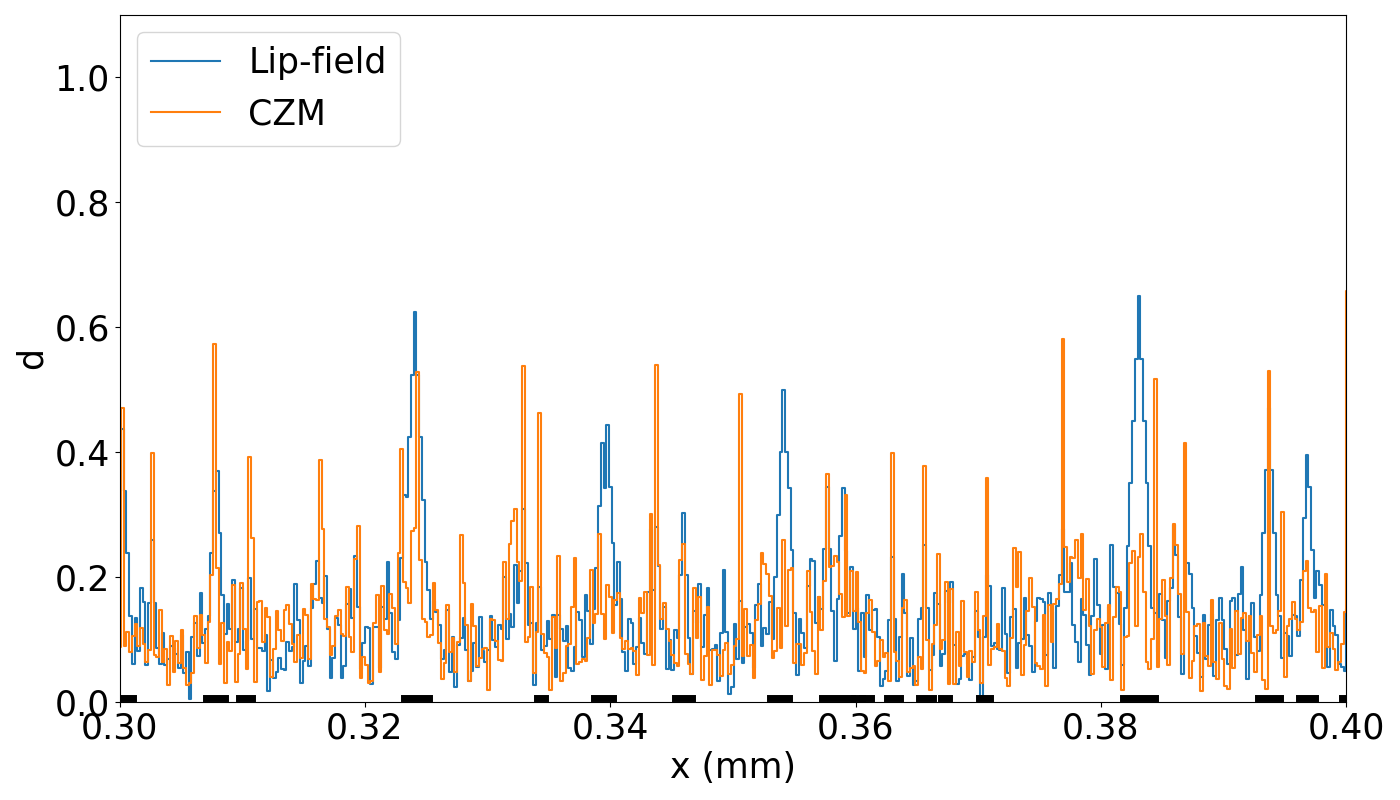}
 \\
 (c) t = 21.196 ns & (d) t = 22.963 ns \\
 \includegraphics[width=7.5cm]{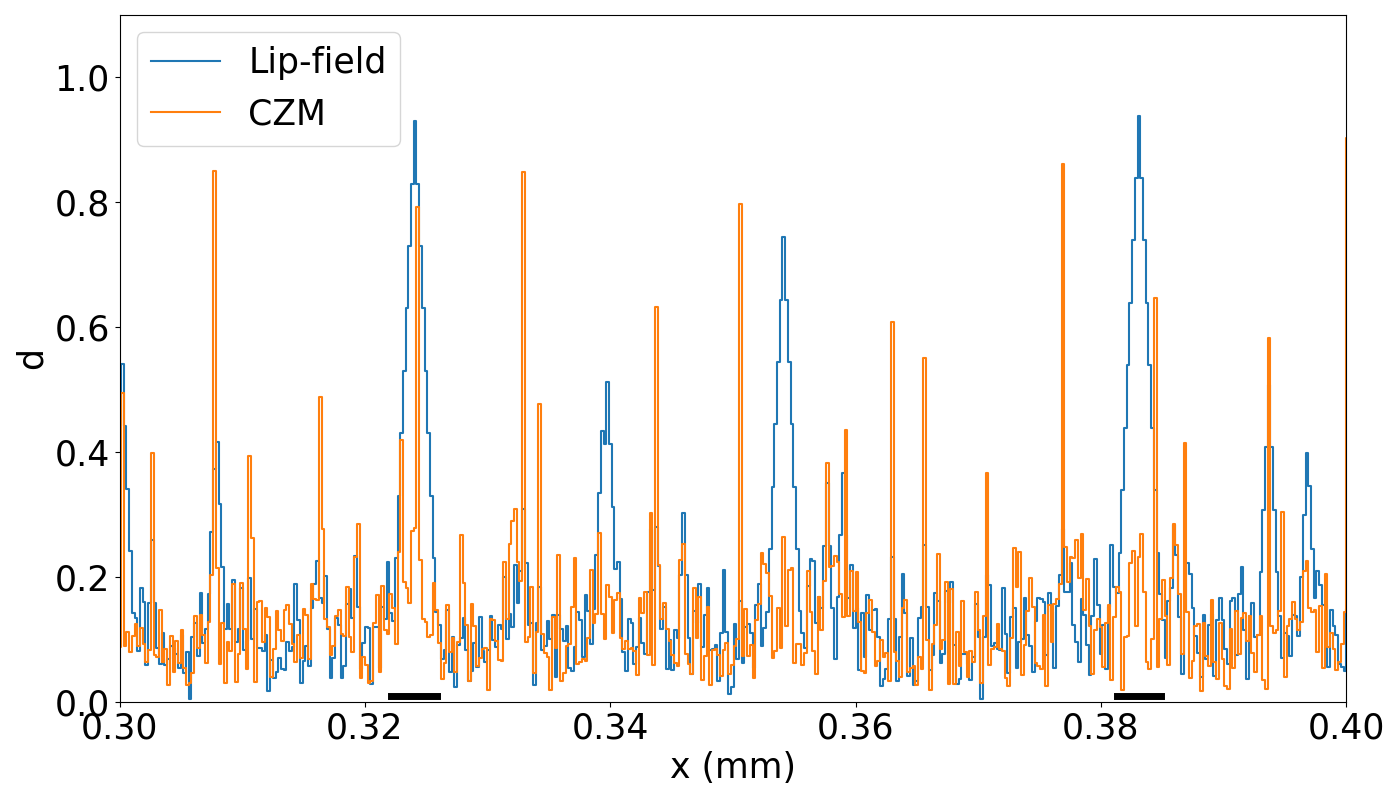}
& 
\includegraphics[width=7.5cm]{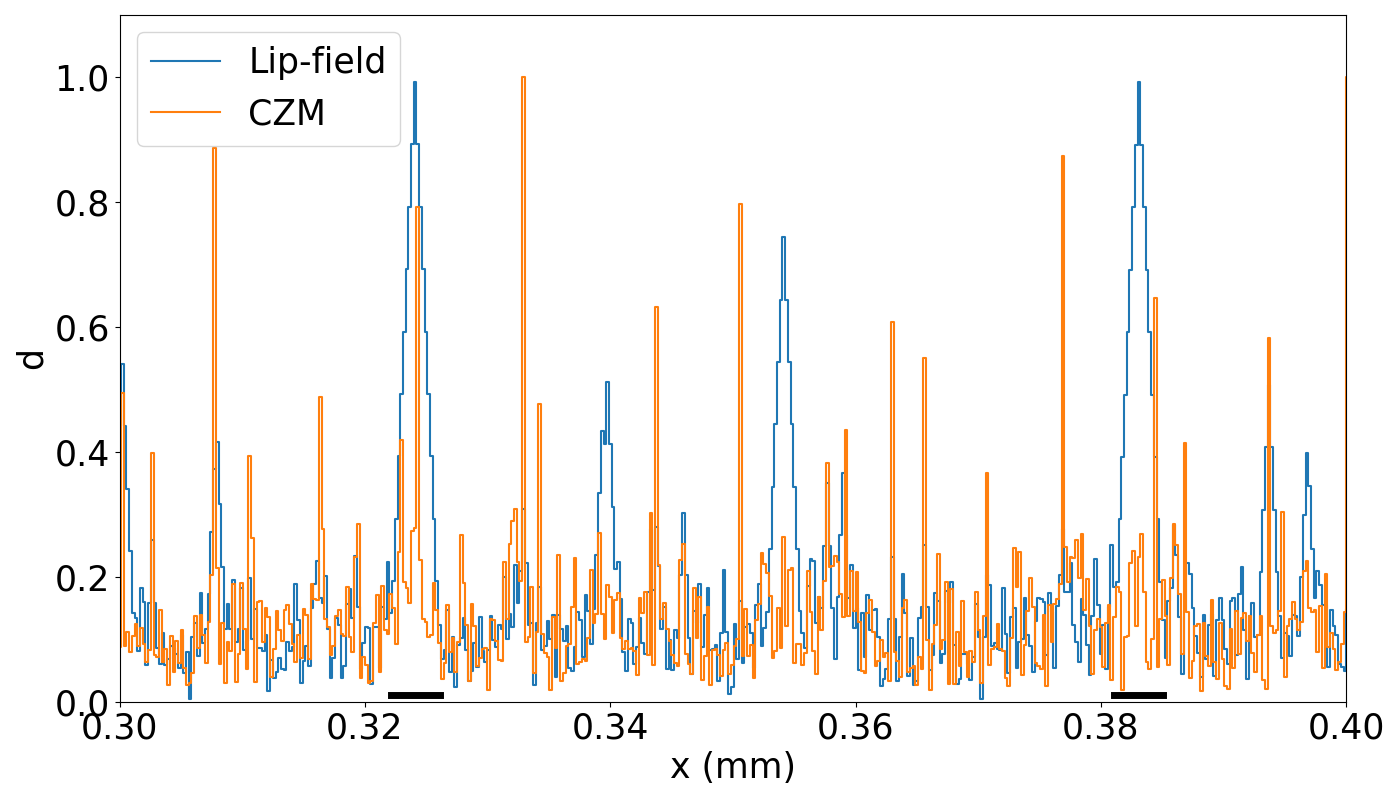}
 \\
 (e) t = 28.262 ns & (f) t = 44.159 ns 
\end{tabular}
\includegraphics[width=7.5cm]{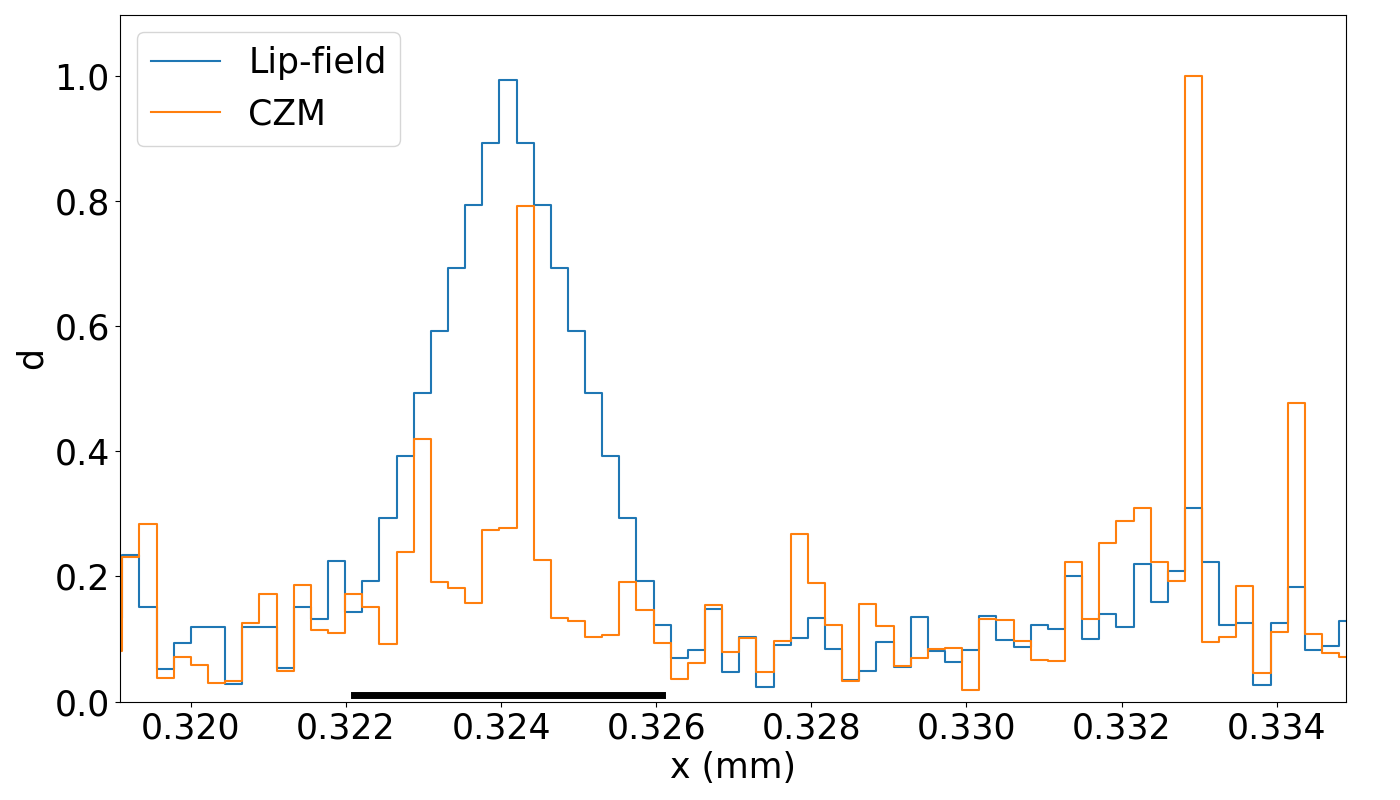} \\
(g) t = 44.159 ns, zoom
\caption{Snapshots of damage field for the Lip-field and CZM in explicit dynamics, for $\epsdoti = 1\cdot 10^5$ s$^{-1}$ and $\helem = \ell / 10$. Black thick horizontal lines indicates where the Lipschitz constraint is active.}
\label{fig:damageComparison}
\end{figure}

In Figure \ref{fig:damageComparison}, snapshots of the damage fields for the Lip-field and the CZM are given at different time steps, for $\epsdoti = 1\cdot 10^5$ s$^{-1}$ and $\helem = \ell / 10$. For the Lip-field, after a first stage where the damage evolves in a purely local way, following the variations of the Young modulus, the Lip-field regularization activates itself in the zones were damage localizes, materialized by the black, thick horizontal lines. The number and size of these zones increase, then decrease until their number corresponds to the final number of cracks. One can observe that the CZM can create cracks with only one finite element. On the other hand, the Lip-field tends to spread damage on zones with a width of about 20 finite elements (see for instance Figure \ref{fig:damageComparison} (g)), which is equal to $2 \ell$.



\begin{figure}
\centering
\begin{tabular}{cc}
 \includegraphics[height=5.5cm]{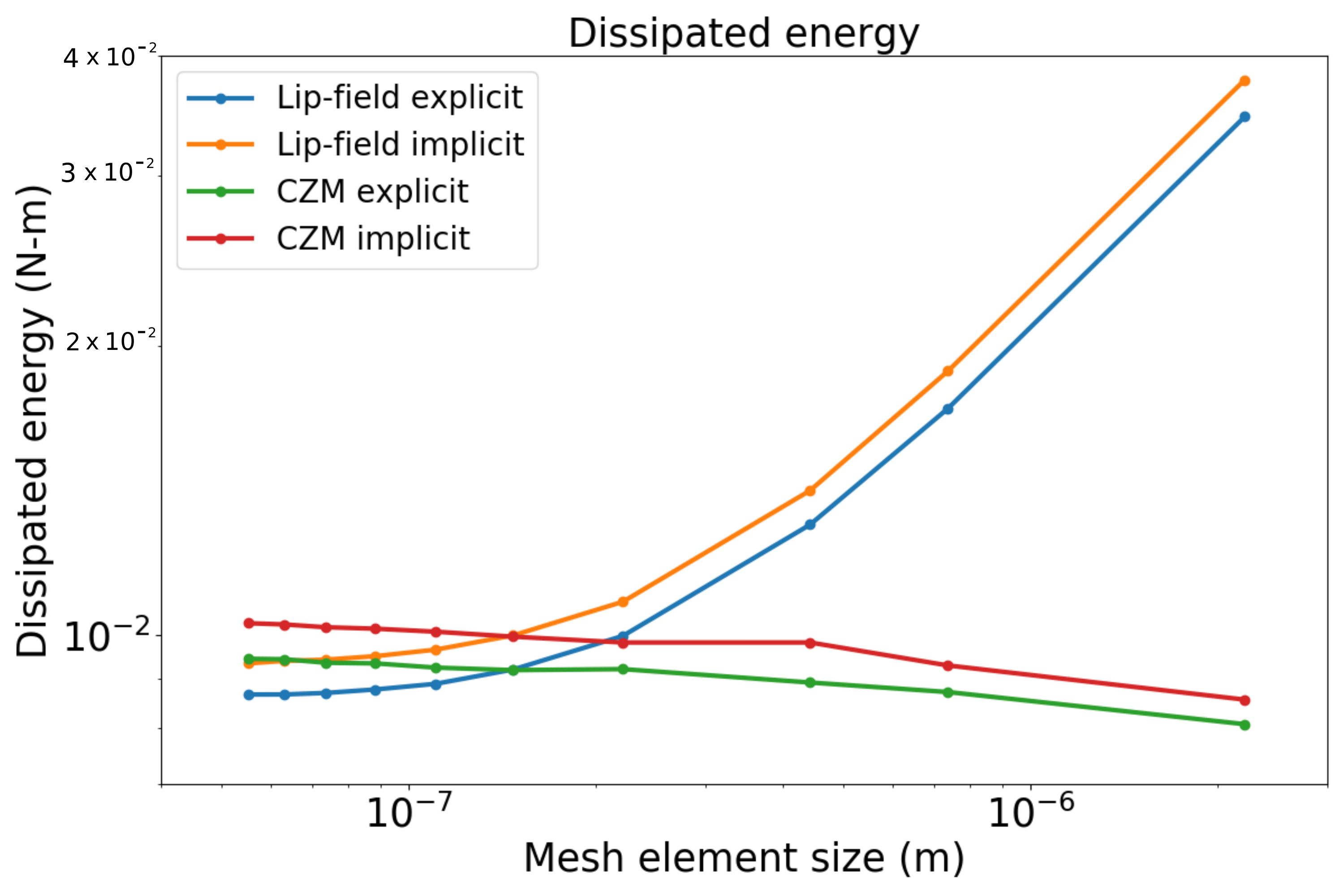}
& 
\includegraphics[height=5.5cm]{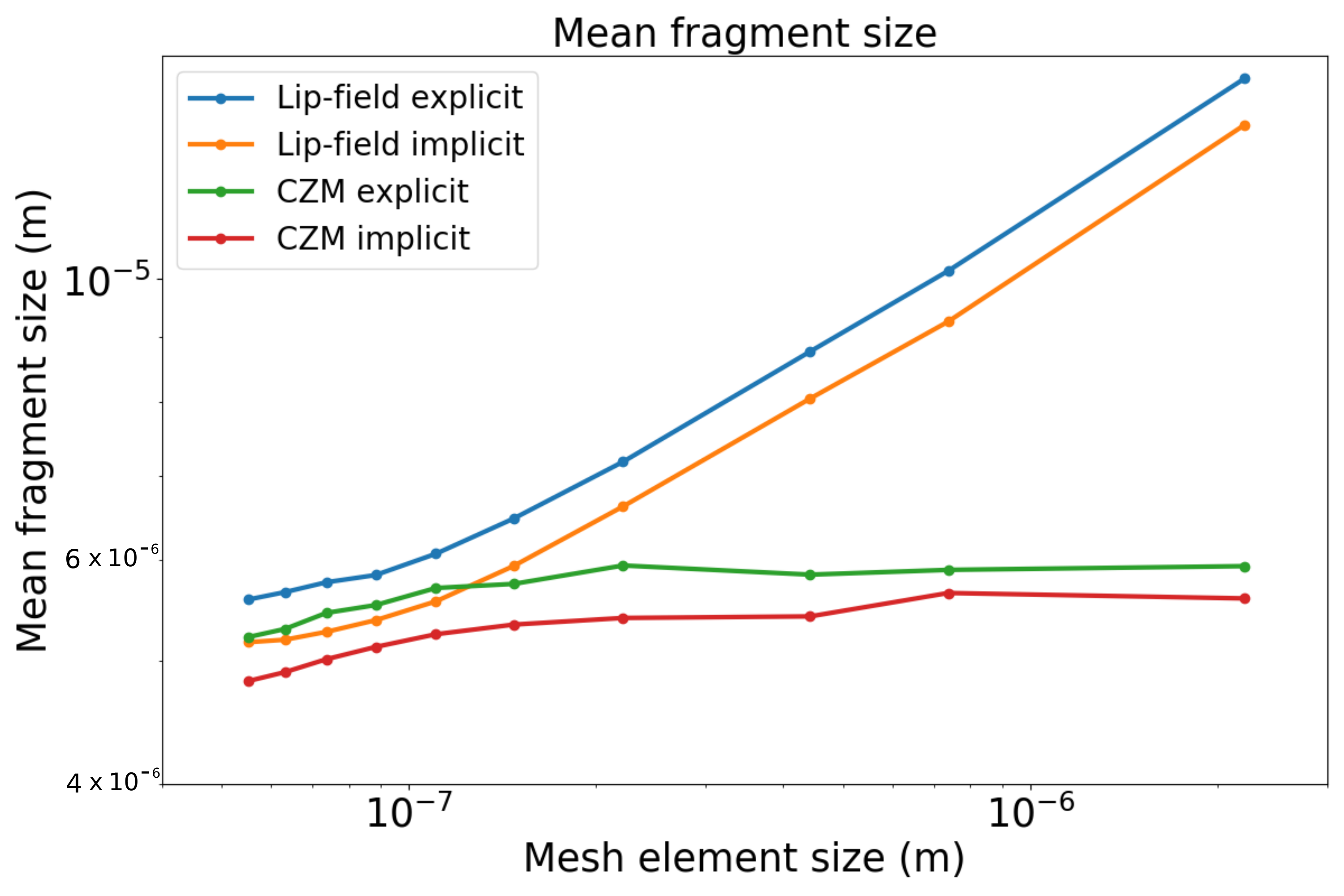}
 \\
(a) & (b)
\end{tabular}
\caption{Lip-field and CZM convergence study for $\epsdoti = 7.5\cdot 10^6$ s$^{-1}$. Dissipated energy is plotted on the left and mean fragment size on the right.
}
\label{fig:convergence}
\end{figure}

Note that in the case of the CZM, even if damage tends to localize in bands of one-element width, the dissipated energy does not tend to zero with mesh refinement, as illustrated on Figure \ref{fig:convergence}. This is of course expected from the crack-band type model used for the CZM. On Figure \ref{fig:convergence} are also plotted the results obtained with the different approaches for different mesh element sizes. On the left is plotted the dissipated energy $D$, computed as:
\begin{equation}
 D = A \int_0^L \Yc h(d) \dint d
\end{equation}

On the right is plotted the mean fragment size, obtained by taking the average value of the distance between two cracks. A crack location is defined as the position of the centroid of any element where $d > 0.98$. Each data point of Figure \ref{fig:convergence} has been obtained by taking the average of 20 runs with different random distributions of the Young modulus.
A first observation is that none of the different computed dissipated energy tends to zero with mesh refinement, which is due to the dependency on the mesh element size of the softening function for the CZM, and due to the Lip-field regularization for the Lip-field. The convergence speed of the CZM is rather fast compared to the Lip-field, with results which are almost independent from the mesh refinement, while the dissipation of the Lip-field stabilizes for a mesh element size of about $\helem \simeq \ell /35$.

Similar behavior was observed in previous studies for a Thick Level Set (TLS) model equivalent to a linear CZM, where fine meshes were necessary to reach convergence \cite{Stershic2017,Le2018}. Comparable studies of dynamic fragmentation using phase-field fracture modeling \cite{Geelen2019,Fischer2019} have also been performed; however, mesh-convergence rate was not reported in 
detail.

A solution was proposed to improve the Lip-field convergence rate, which consisted in introducing cohesive zones prior to having the bulk damage variable going to one, thus avoiding strains going to infinite values in the bulk while preserving the amount of dissipated energy \cite{Le2018}. This solution applied to the Lip-field approach would probably help increasing convergence speed, but is out of the scope of this paper.


\begin{figure}
\centering
\begin{tabular}{cc}
 \includegraphics[height=5.5cm]{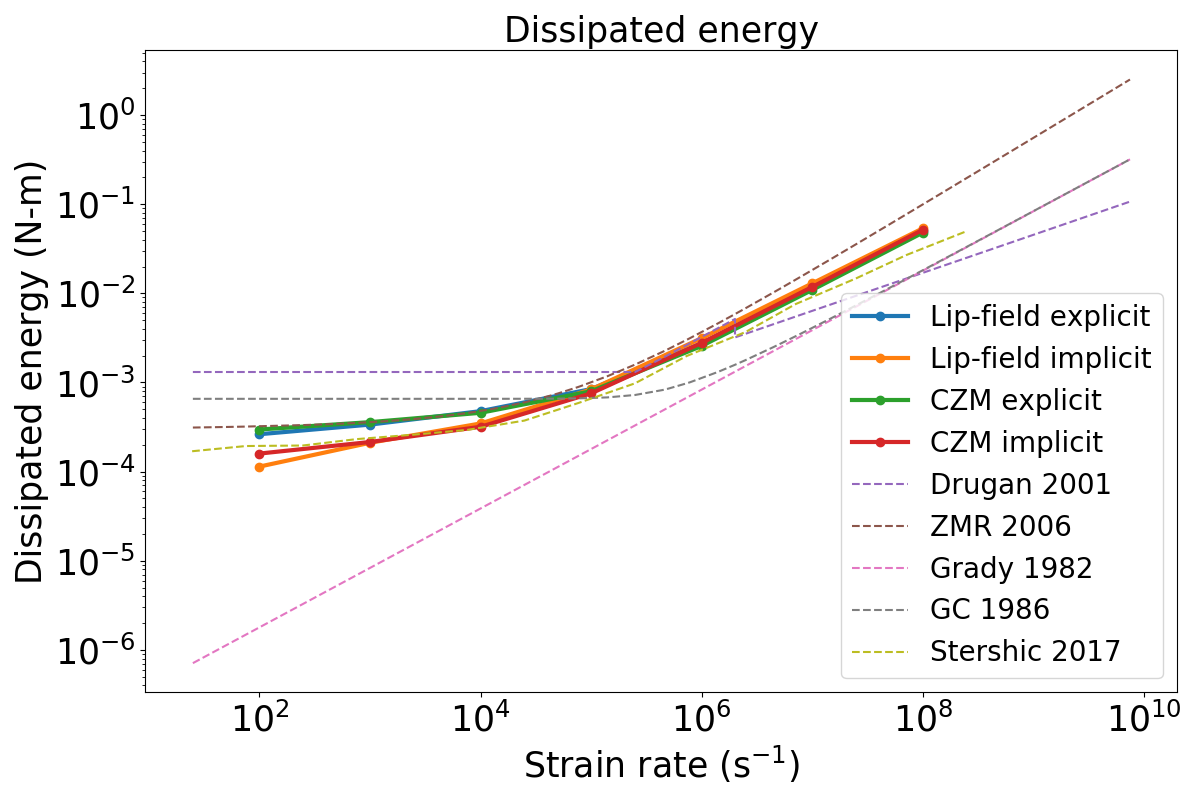}
& 
\includegraphics[height=5.5cm]{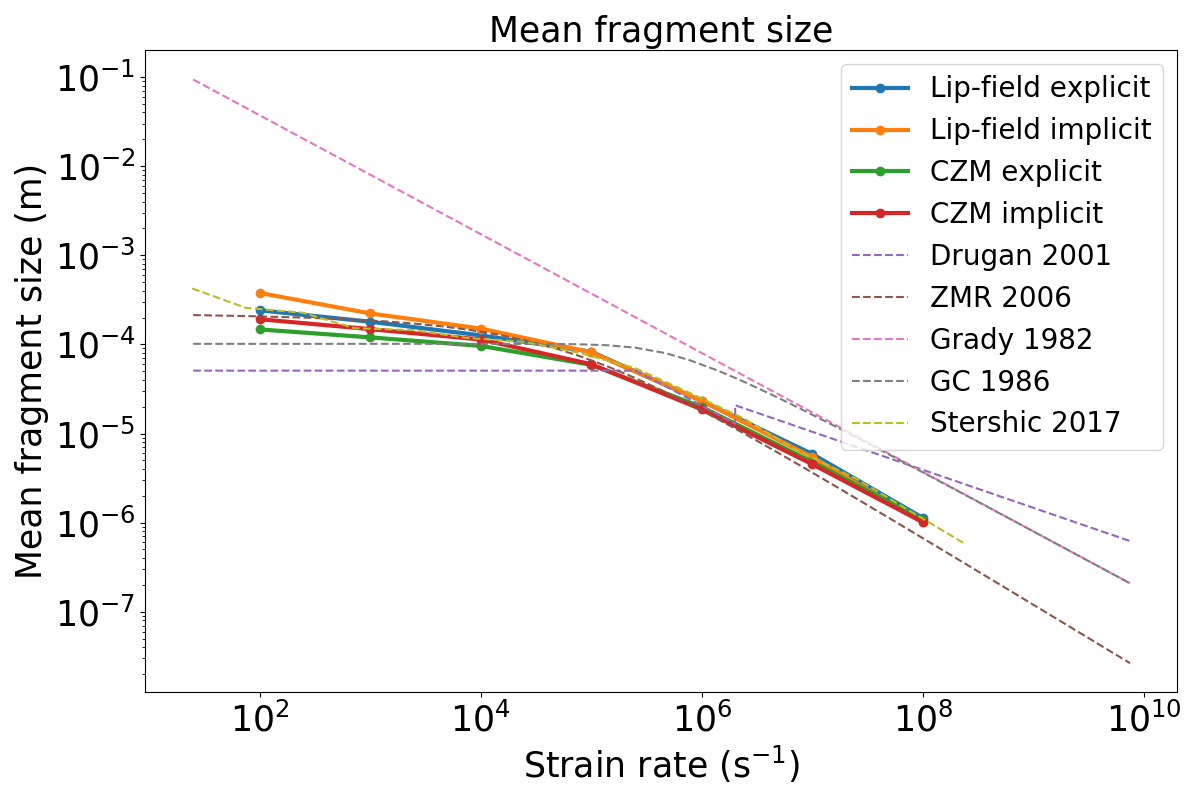}
 \\
(a) & (b)
\end{tabular}
\caption{ Strain rate dependency for $\helem = \lc / 10$. Comparison with Drugan \cite{Drugan2001}, Zhou et al. \cite{Zhou2006}, Grady \cite{Grady1982}, Glenn and Chudnovsky \cite{Glenn1986} and Stershic \cite{Stershic2017}. Dissipated energy is plotted on the left and mean fragment size on the right.}
\label{fig:strainRateDependency}
\end{figure}

The results obtained with a mesh element size of $\helem = \lc / 10$ for different strain rates are compared to different references on Figure \ref{fig:strainRateDependency}. The four tested approaches are able to reproduce the two regimes captured by the different references:  a dissipated energy and mean fragment size that is strongly rate-dependent, while transitioning toward rate-independence at lower strain rates.
The results obtained with the Lip-field match quite well the results obtained with the CZM. For high strain rates, the results obtained with the explicit and implicit approaches are close to each other, while some differences can be observed for low strain rates. However, these differences are within the range of the differences observed between the different references, so they can be considered as acceptable.

\section{Conclusion and future works}
\label{sec:conclusion}

This paper presents the first dynamic application of the Lip-field regularization introduced in \cite{Moes2021}, to the case of 1D dynamic fragmentation. This problem is formulated as an optimization problem, from which we can obtain the classical explicit and implicit dynamic time discretizations. The optimization framework allows to look for the damage variable in a space such that it does not suffer from spurious localization. Parameters of the model are determined to mimic a cohesive zone model (CZM) with a linear traction-separation law, which makes the numerical properties easy to calibrate, knowing the tensile stress and fracture toughness of the material.

We would like to emphasize the simplicity of the proposed approach, that is a simple elastic potential without asymmetry between tension and compression plus a softening function computed to get the proper dissipated energy. The damage variable can be computed using any optimization package by providing the proper Lipschitz constraint. No particular treatment is required to ensure the irreversibility of the damage variable, which is directly taken into account when solving the optimization problem.

For comparison, this paper also introduces a simplified version of a linear CZM consisting in a local damage model, akin to a crack-band model. This allows to run CZM computations without the tedious necessity to implement discrete displacement jumps.  Results exhibited in this paper demonstrate the regularizing effect of the Lip-field approach, with dissipated energies converging to a finite value with finite element mesh refinement.



The numerical results obtained 
for the Lip-field and the CZM over a range of strain rates are compared to several references. 
Both explicit and implicit dynamic simulation results lie well within the variability range of these references and are able to reproduce the same trends.
Along with previous studies \cite{Zhou2006,Molinari2007,Stershic2017}, this work may therefore serve as a benchmark for 1D dynamic fragmentation.
 Because the results of both explicit and implicit simulations converge to the reference range, it is difficult to tell which one is better in terms of accuracy. Further, both approaches seem comparable in terms of mesh convergence rate and give results close to their CZM counterparts.

As the spatial convergence is slow, we expect that there would be a benefit to introduce cohesive zones with displacement jumps prior to the damage variable in the bulk going to one, as demonstrated by \cite{Le2018}.
For 1D problems, the advantage would be to increase the mesh convergence rate. Regarding 2D and 3D problems, placing diffuse Lip-damage around sharp displacement jumps would (i) provide a natural propagation criterion, (ii) help in having crack paths independent from mesh orientation, and (iii) allow modeling of complex crack patterns,
representing a clear advantage over classical cohesive zone models. Future work will serve to verify this expectation.




Finally, further studies could explore the effect of using spatially-correlated random fields for material properties in fragmentation studies. In particular, it would be meaningful to examine the relationship between the material correlation length and the failure length scale as it pertains to fragment size distributions.



\section*{Acknowledgments}

Sandia National Laboratories is a multimission laboratory managed and operated by National Technology \& Engineering Solutions of Sandia, LLC, a wholly owned subsidiary of Honeywell International Inc., for the U.S. Department of Energy's National Nuclear Security Administration under contract DE-NA0003525.

This paper describes objective technical results and analysis. Any subjective views or opinions that might be expressed in the paper do not necessarily represent the views of the U.S. Department of Energy or United States Government.

\appendix


\section*{Appendix: A local damage model equivalent to a linear cohesive zone model}

In this section, we derive the expression of  $h_{\text{CZM}}$ so that the behavior of a 1D element modeled with a local damage model is the same as a 1D element modeled with a linear CZM. The opening at the tip of a 1D element modeled with a CZM with a displacement jump $w$ is:
\begin{equation}
   u = \frac{\helem \sigma}{E} + w
\end{equation}
and for a 1D element with a damage variable $d$:
\begin{equation}
   u = \frac{\helem \sigma}{E \gd(d)}
\end{equation}
so the expression of $w$ is
\begin{equation}
   w = \frac{\helem \sigma}{E} \left( \frac{1-\gd(d)}{\gd(d)} \right)
\end{equation}
By injecting the expression of $w$ above to the linear CZM relation $\sigma = f(w)$, we get
\begin{equation}
 \sigma = \sigc - \lamc I(d)
\end{equation}
where $\lamc = \frac{\sigc \helem}{E \wc}$ and $I(d) = \frac{1-\gd(d)}{\gd(d)}$, which gives the expression of the element stress
\begin{equation}
 \sigma = \frac{\sigc}{1+\lamc I(d)}
\end{equation}
The damage initiation condition gives
\begin{equation}
   \Yc = -\frac{\sigc^2}{2E} \gd'(0)
\end{equation}
and the damage growth condition is
\begin{equation}
   - \frac{E \gd'(d)}{2}  \epsilon^2 = - \frac{ \gd'(d)}{2 E} \sigma^2 = \Yc 
H_{\text{CZM}}(d)
   \label{eq:czmdamagegrowth}
\end{equation}

By combining the above equations, the expression of $H_{\text{CZM}}$ is
\begin{equation}
 H_{\text{CZM}}(d) = - \frac{\gd'(d)}{\left((1-\lamc) \gd(d) + \lamc\right)^2}
\end{equation}
and the expression of $h_{\text{CZM}}$ is
\begin{equation}
 h_{\text{CZM}}(d)  = \frac{1}{(1-\lamc)} \left( \frac{1}{(1-\lamc) \gd(d) + \lamc} - 1 \right)
\end{equation}

\newpage

\bibliography{refs}

\begin{thebibliography}{10}

\bibitem{Moes2021}
N.~Mo{\"{e}}s and N.~Chevaugeon.
\newblock {Lipschitz regularization for softening material models: the
  Lip-field approach}.
\newblock {\em Comptes Rendus. M{\'{e}}canique}, 349(2):415--434, 2021.

\bibitem{Chevaugeon2021}
N.~Chevaugeon and N.~Mo\"es.
\newblock Lipschitz regularization for fracture: the {L}ip-field approach.
\newblock {\em arXiv preprint arXiv:2111.04771}, 2021.

\bibitem{Denoual2000}
C.~Denoual and F.~Hild.
\newblock A damage model for the dynamic fragmentation of brittle solids.
\newblock {\em Computer Methods in Applied Mechanics and Engineering},
  183(3-4):247--258, 2000.

\bibitem{Mott1947}
N.F. Mott.
\newblock Fragmentation of shell cases.
\newblock {\em Proceedings of the Royal Society of London. Series A.
  Mathematical and Physical Sciences}, 189(1018):300--308, 1947.

\bibitem{Grady1982}
D.E. Grady.
\newblock Local inertial effects in dynamic fragmentation.
\newblock {\em Journal of Applied Physics}, 53(1):322--325, 1982.

\bibitem{Glenn1986}
L.A. Glenn and A.~Chudnovsky.
\newblock Strain-energy effects on dynamic fragmentation.
\newblock {\em Journal of Applied Physics}, 59(4):1379--1380, 1986.

\bibitem{Drugan2001}
W.~J. Drugan.
\newblock {Dynamic fragmentation of brittle materials: Analytical
  mechanics-based models}.
\newblock {\em Journal of the Mechanics and Physics of Solids},
  49(6):1181--1208, 2001.

\bibitem{Miller1999}
O.~Miller, L.B. Freund, and A.~Needleman.
\newblock {Modeling and simulation of dynamic fragmentation in brittle
  materials}.
\newblock {\em International Journal of Fracture}, 96(2):101--125, 1999.

\bibitem{Pandolfi1999}
A.~Pandolfi, P.~Krysl, and M.~Ortiz.
\newblock Finite element simulation of ring expansion and fragmentation: the
  capturing of length and time scales through cohesive models of fracture.
\newblock {\em International Journal of Fracture}, 95(1):279--297, 1999.

\bibitem{Zhou2006}
F.~Zhou, J.-F. Molinari, and K.T. Ramesh.
\newblock {Effects of material properties on the fragmentation of brittle
  materials}.
\newblock {\em International Journal of Fracture}, 139(2):169--196, 2006.

\bibitem{Pearson1990}
J.~Pearson.
\newblock A fragmentation model for cylindrical warheads.
\newblock Technical report, Naval Weapons Center, China Lake, CA, 1990.

\bibitem{Tonge2016}
A.L. Tonge and K.T. Ramesh.
\newblock Multi-scale defect interactions in high-rate failure of brittle
  materials, {P}art {II}: Application to design of protection materials.
\newblock {\em Journal of the Mechanics and Physics of Solids}, 86:237--258,
  2016.

\bibitem{Selvadurai2009}
A.P.S. Selvadurai.
\newblock Fragmentation of ice sheets during impact.
\newblock {\em Computer Modeling in Engineering and Science}, 52:259--277,
  2009.

\bibitem{Aastrom2019}
J.A. {\AA}str{\"o}m and D.I. Benn.
\newblock Effective rheology across the fragmentation transition for sea ice
  and ice shelves.
\newblock {\em Geophysical Research Letters}, 46(22):13099--13106, 2019.

\bibitem{Bishop2016}
J.E. Bishop, M.J. Martinez, and P.~Newell.
\newblock Simulating fragmentation and fluid-induced fracture in disordered
  media using random finite-element meshes.
\newblock {\em International Journal for Multiscale Computational Engineering},
  14(4), 2016.

\bibitem{Hu2020}
T.~Hu, J.~Guilleminot, and J.E. Dolbow.
\newblock A phase-field model of fracture with frictionless contact and random
  fracture properties: Application to thin-film fracture and soil desiccation.
\newblock {\em Computer Methods in Applied Mechanics and Engineering},
  368:113106, 2020.

\bibitem{Jiang2020}
W.~Jiang, B.W. Spencer, and J.E. Dolbow.
\newblock Ceramic nuclear fuel fracture modeling with the extended finite
  element method.
\newblock {\em Engineering Fracture Mechanics}, 223:106713, 2020.

\bibitem{Dugdale1960}
D.S. Dugdale.
\newblock {Yielding of steel sheets containing slits}.
\newblock {\em Journal of the Mechanics and Physics of Solids}, 8:100--104,
  1960.

\bibitem{Baranblatt1961}
G.I. Baranblatt.
\newblock {The mathematical theory of equilibrium cracks formed by brittle
  fracture}.
\newblock {\em Zh. Prkl. Mekh. Tekh. Fiz}, 25(4):3--56, 1961.

\bibitem{hilleborg1976}
A.~Hillerborg, M.~Mod{\'e}er, and P-E. Petersson.
\newblock {Analysis of crack formation and crack growth in concrete by means of
  fracture mechanics and finite elements}.
\newblock {\em Cement and Concrete Research}, 6:773--782, 1976.

\bibitem{Park2011}
K.~Park and G.H. Paulino.
\newblock {Cohesive zone models: A critical review of traction-separation
  relationships across fracture surfaces}.
\newblock {\em Applied Mechanics Reviews}, 64(6), 2011.

\bibitem{Moes2002}
N.~Mo{\"{e}}s and T.~Belytschko.
\newblock {Extended finite element method for cohesive crack growth}.
\newblock {\em Engineering Fracture Mechanics}, 69(7):813--833, May 2002.

\bibitem{Pramanik2015}
R.~Pramanik and D.~Deb.
\newblock Implementation of smoothed particle hydrodynamics for detonation of
  explosive with application to rock fragmentation.
\newblock {\em Rock Mechanics and Rock Engineering}, 48(4):1683--1698, 2015.

\bibitem{Li2015}
B.~Li, A.~Pandolfi, and M.~Ortiz.
\newblock Material-point erosion simulation of dynamic fragmentation of metals.
\newblock {\em Mechanics of Materials}, 80:288--297, 2015.

\bibitem{Kun1996}
F.~Kun and H.J. Herrmann.
\newblock A study of fragmentation processes using a discrete element method.
\newblock {\em Computer Methods in Applied Mechanics and Engineering},
  138(1-4):3--18, 1996.

\bibitem{Lai2015}
X.~Lai, B.~Ren, H.~Fan, S.~Li, C.T. Wu, R.A. Regueiro, and L.~Liu.
\newblock Peridynamics simulations of geomaterial fragmentation by impulse
  loads.
\newblock {\em International Journal for Numerical and Analytical Methods in
  Geomechanics}, 39(12):1304--1330, 2015.

\bibitem{Kachanov1958}
L.~Kachanov.
\newblock {Rupture Time Under Creep Conditions}.
\newblock {\em Otdelenie tekhnicheskich nauk}, pages 26--31, 1958.

\bibitem{chaboche1988}
J.-L. Chaboche and J.~Lemaitre.
\newblock {\em {M{\'{e}}canique des mat{\'{e}}riaux solides}}.
\newblock Paris, Bordas, Paris, dunod edition, 1988.

\bibitem{Bazant1976}
Z.P. Ba{\v{z}}ant.
\newblock {Instability, Ductility and Size-effect in Strain-Softening
  Concrete}.
\newblock {\em Journal Of The Engineering Mechanics Division}, 102(2):331--343,
  1976.

\bibitem{Cosserat1909}
E.~Cosserat and F.~Cosserat.
\newblock {\em {Th{\'{e}}orie des corps d{\'{e}}formables}}.
\newblock Herman, Paris, 1909.

\bibitem{Chambon1998}
R.~Chambon, D.~Caillerie, and N.~{El Hassan}.
\newblock {One-dimensional localisation studied with a second grade model}.
\newblock {\em European Journal of Mechanics - A/Solids}, 17(4):637--656, 1998.

\bibitem{Pijaudier1987}
G.~Pijaudier-Cabot and Z.P. Ba{\v{z}}ant.
\newblock {Non local damage theory}.
\newblock {\em Journal of Engineering Mechanics}, 113(10):1512--1533, 1987.

\bibitem{Lasry1988}
D.~Lasry and T.~Belytschko.
\newblock {Localization limiters in transient problems}.
\newblock {\em International Journal of Solids and Structures}, 24(6):581--597,
  1988.

\bibitem{Peerlings1996}
R.H.J. Peerlings, R.~de~Borst, W.~A.~M. Brekelmans, and J.~H.~P. {De Vree}.
\newblock {Gradient Enhanced Damage for Quasi-Brittle Materials}.
\newblock {\em International Journal for Numerical Methods in Engineering},
  39(19):3391--3403, oct 1996.

\bibitem{Jirasek1998}
M.~Jir\'asek.
\newblock Nonlocal models for damage and fracture: comparison of approaches.
\newblock {\em International Journal of Solids and Structures},
  35(31-32):4133--4145, 1998.

\bibitem{giry2011b}
C.~Giry, F.~Dufour, and J.~Mazars.
\newblock {Stress-based nonlocal damage model}.
\newblock {\em International Journal of Solids and Structures}, 48:3431--3443,
  2011.

\bibitem{Needleman1988}
A.~Needleman.
\newblock {Material rate dependence and mesh sensitivity in localization
  problems}.
\newblock {\em Computer Methods in Applied Mechanics and Engineering},
  67(1):69--85, 1988.

\bibitem{Dube1996}
J.F. Dub{\'{e}}, G.~Pijaudier-Cabot, and C.~{La Borderie}.
\newblock {Rate dependent damage model for concrete in dynamics}.
\newblock {\em Journal of Engineering Mechanics}, 122(10):939--947, 1996.

\bibitem{Allix1997}
O.~Allix and J.F De{\"{u}}.
\newblock {Delayed-damage modelling for fracture prediction of laminated
  composites under dynamic loading}.
\newblock {\em Engineering transactions}, 45(1):29--46, 1997.

\bibitem{Suffis2003}
A.~Suffis, T.A.A. Lubrecht, and A.~Combescure.
\newblock {Damage model with delay effect analytical and numerical studies of
  the evolution of the characteristic damage length}.
\newblock {\em International Journal of Solids and Structures},
  40(13-14):3463--3476, 2003.

\bibitem{Desmorat2010}
R.~Desmorat, M.~Chambart, F.~Gatuingt, and D.~Guilbaud.
\newblock {Delay-active damage versus non-local enhancement for anisotropic
  damage dynamics computations with alternated loading}.
\newblock {\em Engineering Fracture Mechanics}, 77(12):2294--2315, 2010.

\bibitem{Zghal2020}
J.~Zghal and N.~Mo\"es.
\newblock Analysis of the delayed damage model for three one-dimensional
  loading scenarii.
\newblock {\em Comptes Rendus. Physique}, 21(6):527--537, 2020.

\bibitem{Karma2001}
A.~Karma, D.~Kessler, and H.~Levine.
\newblock {Phase-Field Model of Mode III Dynamic Fracture}.
\newblock {\em Physical Review Letters}, 87(4):045501, 2001.

\bibitem{Miehe2010}
C.~Miehe, F.~Welschinger, and M.~Hofacker.
\newblock {Thermodynamically consistent phase-field models of fracture:
  Variational principles and multi-field FE implementations}.
\newblock {\em International Journal for Numerical Methods in Engineering},
  83(10):1273--1311, 2010.

\bibitem{Francfort1998}
G.~A. Francfort and J-J. Marigo.
\newblock {Revisiting brittle fracture as an energy minimization problem}.
\newblock {\em Journal of the Mechanics and Physics of Solids},
  46(8):1319--1342, 1998.

\bibitem{Verhoosel2013}
C.V. Verhoosel and R.~de~Borst.
\newblock A phase-field model for cohesive fracture.
\newblock {\em International Journal for Numerical Methods in Engineering},
  96(1):43--62, 2013.

\bibitem{Wu2018}
J-Y. Wu.
\newblock {A geometrically regularized gradient-damage model with energetic
  equivalence}.
\newblock {\em Computer Methods in Applied Mechanics and Engineering},
  328:612--637, 2018.

\bibitem{Talamini2021}
B.~Talamini, M.R. Tupek, A.J. Stershic, T.~Hu, J.W. Foulk~III, J.T. Ostien, and
  J.E. Dolbow.
\newblock Attaining regularization length insensitivity in phase-field models
  of ductile failure.
\newblock {\em Computer Methods in Applied Mechanics and Engineering},
  384:113936, 2021.

\bibitem{Borden2012}
M.~J. Borden, C.V. Verhoosel, M.~A. Scott, T.J.R. Hughes, and C.M. Landis.
\newblock {A phase-field description of dynamic brittle fracture}.
\newblock {\em Computer Methods in Applied Mechanics and Engineering},
  217:77--95, 2012.

\bibitem{Ren2019}
H.~L. Ren, X.~Y. Zhuang, C.~Anitescu, and T.~Rabczuk.
\newblock {An explicit phase field method for brittle dynamic fracture}.
\newblock {\em Computers and Structures}, 217:45--56, 2019.

\bibitem{Geelen2019}
R.J.M. Geelen, Y.~Liu, T.~Hu, M.R. Tupek, and J.E. Dolbow.
\newblock {A phase-field formulation for dynamic cohesive fracture}.
\newblock {\em Computer Methods in Applied Mechanics and Engineering},
  348:680--711, 2019.

\bibitem{Fischer2019}
A.G. Fischer and J.-J. Marigo.
\newblock Gradient damage models applied to dynamic fragmentation of brittle
  materials.
\newblock {\em International Journal of Fracture}, 220(2):143--165, 2019.

\bibitem{moes2011}
N.~Mo{\"{e}}s, C.~Stolz, P.E. Bernard, and N.~Chevaugeon.
\newblock {A level set based model for damage growth : the thick level set
  approach}.
\newblock {\em International Journal for Numerical Methods in Engineering},
  86(3):358--380, 2011.

\bibitem{Stolz2012b}
C.~Stolz and N.~Mo{\"{e}}s.
\newblock {A new model of damage: a moving thick layer approach}.
\newblock {\em International Journal of Fracture}, 174:49--60, 2012.

\bibitem{Bernard2012}
P.E. Bernard, N.~Mo{\"{e}}s, and N.~Chevaugeon.
\newblock {Damage growth modeling using the Thick Level Set (TLS) approach:
  Efficient discretization for quasi-static loadings}.
\newblock {\em Computer Methods in Applied Mechanics and Engineering},
  233-236:11--27, August 2012.

\bibitem{salzman2016}
A.~Salzman, N.~Mo{\"{e}}s, and N.~Chevaugeon.
\newblock {On use of the Thick Level Set method in 3D quasi-static crack
  simulation of quasi-brittle material}.
\newblock {\em International Journal of Fracture}, 202(1):21--49, 2016.

\bibitem{Moreau2015}
K.~Moreau, N.~Mo{\"{e}}s, D.~Picart, and L.~Stainier.
\newblock {Explicit dynamics with a non-local damage model using the thick
  level set approach}.
\newblock {\em International Journal for Numerical Methods in Engineering},
  102(3-4):808--838, 2015.

\bibitem{Stershic2017}
A.~J. Stershic, J.E Dolbow, and N.~Mo{\"{e}}s.
\newblock {The Thick Level-Set model for dynamic fragmentation}.
\newblock {\em Engineering Fracture Mechanics}, 172:39--60, 2017.

\bibitem{ParrillaGomez2015}
A.~{Parrilla G{\'{o}}mez}, N.~Mo{\"{e}}s, and C.~Stolz.
\newblock {Comparison between Thick Level Set (TLS) and cohesive zone models}.
\newblock {\em Advanced Modeling and Simulation in Engineering Sciences},
  2(1):1--22, 2015.

\bibitem{Bazant1982}
Z.P. Ba{\v{z}}ant.
\newblock {Crack Band Model for Fracture of Geomaterials.}
\newblock In {\em Proceedings of the Fourth International Conference on
  Numerical Methods in Geomechanics}, volume III, pages 1137--1152, 1982.

\bibitem{bazant1983}
Z.P. Ba{\v{z}}ant and B.H. Oh.
\newblock {Crack band theory for fracture of concrete}.
\newblock {\em Materials and Structures}, 16:155--177, 1983.

\bibitem{Molinari2007}
J.-F. Molinari, G.~Gazonas, R.~Raghupathy, A.~Rusinek, and F.~Zhou.
\newblock The cohesive element approach to dynamic fragmentation: the question
  of energy convergence.
\newblock {\em International Journal for Numerical Methods in Engineering},
  69(3):484--503, 2007.

\bibitem{Le2018}
B.~L{\'{e}}, N.~Mo{\"{e}}s, and G.~Legrain.
\newblock {Coupling damage and cohesive zone models with the Thick Level Set
  approach to fracture }.
\newblock {\em Engineering Fracture Mechanics}, 193:214--247, 2018.

\end{thebibliography}
\bibliographystyle{unsrt}

\end{document}